\documentclass[journal]{IEEEtran}

\usepackage{times}
\usepackage{epsfig}
\usepackage{graphicx}
\usepackage{amsmath}
\usepackage{amssymb}
\usepackage{enumitem}
\usepackage{array}%需要该宏包
\usepackage{multirow}
\usepackage{graphicx}
\usepackage{subfigure}
\usepackage{amsthm,amsmath,amssymb}
\usepackage{mathrsfs}
\usepackage{booktabs}
\usepackage{cite}
\usepackage{color}
\usepackage{pifont}
\usepackage{lipsum}
\usepackage{bm}
\usepackage[acronym]{glossaries}
\newcommand{\newac}{\newacronym}
\newcommand{\ac}{\gls}
\newcommand{\Ac}{\Gls}
\newcommand{\acpl}{\glspl}

\makeglossaries
\newac{hmm}{HMM}{hidden Markov model}
\newac{wcl}{WCL}{weighted central localization}
\newac{dft}{DFT}{discrete Fourier transform}
\newac{csi}{CSI}{channel state information}
\newac{tdoa}{TDoA}{time-difference-of-arrival}
\newac{gnss}{GNSS}{global navigation satellite system}
\newac{imu}{IMU}{inertial measurement unit}
\newac{bs}{BS}{base station}
\newac{wrt}{w.r.t.}{with respect to}
\newac{pdf}{PDF}{probability density function}
\newac{rss}{RSS}{received signal strength}
\newac{rfid}{RFID}{radio frequency identification detectors}
\newac{gps}{GPS}{global positioning system}
\newac{mems}{MEMS}{microelectromechanical system}
\newac{ins}{INS}{inertial navigation system}
\newac{lidar}{Lidar}{light detection and ranging}
\newac{radar}{Radar}{radio detection and ranging}
\newac{rtk}{RTK}{real-time kinematic positioning}
\newac{v2i}{V2I}{vehicular communication systems}
\newac{los}{LOS}{line of sight}
\newac{vtr}{VTR}{vehicle trajectory reconstruction}

\usepackage[table]{xcolor}
\usepackage{float}
\floatstyle{ruled}
\newfloat{algorithm}{tbp}{loa}
\providecommand{\algorithmname}{Algorithm}
\floatname{algorithm}{\protect\algorithmname}

\begin{document}
	
	\title{Trajectory Map-Matching in Urban Road Networks\\ Based on RSS Measurements}

	\author{
		Zheng~Xing and Weibing~Zhao
		
%		\thanks{\indent Manuscript received 11 November 2023; revised 17 May 2024, 8 October 2024, and 17 December 2024; accepted January 31, 2025. (Corresponding author: Weibing Zhao.)
%		}
		
%		\thanks{\indent Zheng~Xing and Weibing~Zhao are with Guangdong Laboratory of Machine Perception and Intelligent Computing, Shenzhen MSU-BIT University, Shenzhen, P.R. China. Zheng Xing is also with Future Network of Intelligence Institute, School of Science and Engineering, The Chinese University of Hong Kong, Shenzhen, China. (E-mail: zhengxing@link.cuhk.edu.cn; weibingzhao@smbu.edu.cn)}

	} %
	
%	\markboth{IEEE TRANSACTIONS ON INTELLIGENT TRANSPORTATION SYSTEMS}%
%	{aa}
	
	\maketitle
	
	\begin{abstract}
		
		The widespread deployment of wireless communication networks has catalyzed significant advancements in utilizing signal channs to address real-world challenges, such as \ac{vtr}, drone trajectory planning, and network optimization. Existing methods primarily utilize \ac{tdoa} measurements for vehicle localization. However, these methods require specialized decoding receivers capable of deciphering communication protocols, leading to increased application costs.
		\Ac{rss}, a measure of wireless signal strength, can be recorded by any standard communication device, thus allowing \ac{rss}-based \ac{vtr} to benefit from cost-effectiveness. Nevertheless, the inherently noisy and sporadic nature of \ac{rss} poses significant challenges for accurately reconstructing vehicle trajectories.
		This paper aims to utilize RSS measurements to reconstruct vehicle trajectories within a road network. We constrain the trajectories to comply with signal propagation rules and vehicle mobility constraints, thereby mitigating the impact of the noisy and sporadic nature of RSS data on the accuracy of trajectory reconstruction.
		The primary challenge involves exploiting latent spatial-temporal correlations within the noisy and sporadic \ac{rss} data while navigating the complex road network.
		To overcome these challenges, we develop an \ac{hmm}-based \ac{rss} embedding (HRE) technique that utilizes alternating optimization to search for the vehicle trajectory based on \ac{rss} measurements. This model effectively captures the spatial-temporal relationships among \ac{rss} measurements, while a road graph model ensures compliance with network pathways. Additionally, we introduce a maximum speed-constrained rough trajectory estimation (MSR) method to effectively guide the proposed alternating optimization procedure, ensuring that the proposed HRE method rapidly converges to a favorable local solution.
		The proposed method is validated using real \ac{rss} measurements from 5G NR networks in Chengdu and Shenzhen, China. The experimental results demonstrate that the proposed approach significantly outperforms state-of-the-art methods, even with limited \ac{rss} data.
	\end{abstract}


	\IEEEpeerreviewmaketitle
	
	\glsreset{vtr}\glsreset{hmm}\glsreset{rss}\glsreset{tdoa}

	\section{Introduction}

	\Ac{vtr} has garnered considerable attention from both practitioners and researchers due to its significant promise in enhancing vehicle management and surveillance, estimating vehicle flow, controlling traffic signals, and planning urban resources. Despite its potential, the widespread application of \ac{vtr} is hindered by high barriers to data sharing and constraints related to devices and environments, often making vehicle traces inaccessible \cite{Whi:J22}.
	
	Existing \ac{vtr} methods can be broadly categorized into four primary approaches. The \ac{gnss}-aided localization method, the most commonly employed, utilizes position information from \ac{gnss} at each moment to localize and reconstruct vehicle trajectories. This method achieves errors of less than 6 meters under \ac{los} conditions, as shown in Table~\ref{tab:comparison}. When combined with sensors such as \ac{rtk}, \ac{lidar}, \ac{radar}, \ac{ins}, sonar, and cameras \cite{lu2019l3,hata2015feature,chen2022vehicle,xing2024calibration}, it can reach sub-centimeter-level accuracy. However, \ac{gnss}-based tools often fall short in capturing comprehensive traffic flow data essential for subsequent applications.

	\begin{table*}
		%		\arrayrulecolor{blue}
		\caption{Comparisons of different localization methods }
		\centering
		\begin{tabular}{>{\raggedright}p{3cm}>{\centering}p{2cm}>{\raggedright}p{4cm}>{\raggedright}p{6cm}c} 
			\toprule[1.5pt] 
			
			{\textit{Works}} & {\textit{Error}} & {\textit{Required equipment}}  & {\textit{Application environment}}& \textit{Dataset}\\
			\bottomrule[1pt]  
			\rowcolor{gray!50}\multicolumn{5}{l}{\textit{\ac{gnss}-aided localization method}}\tabularnewline
			
			\hline
			
			\rowcolor{gray!20}Xu et al. \cite{xu2018enhancing} & 2.32 m & \ac{mems}-based \ac{ins},\ac{gps} & Covered by satellite signals & Real \tabularnewline

			Lu et al. \cite{lu2019l3} & 1.59 cm$\sim$5.50 cm & \ac{lidar}, \ac{imu}, \ac{ins},
			\ac{rtk}-based \ac{gnss} & Covered by satellite signals & Real \tabularnewline

			\rowcolor{gray!20} Hata et al. \cite{hata2015feature} & 14 cm$\sim$27 cm & \ac{ins}, \ac{gps}, \ac{imu}
			, \ac{lidar} & Covered by satellite signals & Real \tabularnewline

			Chen et al. \cite{chen2022vehicle} & 5.25 m & \ac{gps}, stereovision camera, \ac{lidar},
			\ac{radar}, sonar & Covered by satellite signals, connected and automated vehicle environment & Real \tabularnewline
			
			\bottomrule[1pt] 
			\rowcolor{gray!50}\multicolumn{5}{l}{\textit{GNSS-aided map-matching  method}}\tabularnewline
			\hline
			\rowcolor{gray!20} Paul et al. \cite{newson2009hidden} & $<10$ m & GPS
			& Covered by satellite signals & Real\tabularnewline
			
			Yin et al. \cite{lou2009map} & $<10$ m & GPS
			& Covered by satellite signals & Real\tabularnewline
			
			\rowcolor{gray!20} Hsueh et al. \cite{hsueh2018map} &$<10$ m& GPS
			& Covered by satellite signals & Real\tabularnewline
			
			Jin et al. \cite{jin2022transformer} &$<10$ m& GPS
			& Covered by satellite signals & Real\tabularnewline
			
			\rowcolor{gray!20} Hu et al. \cite{hu2023amm} & $<10$ m & GPS
			& Covered by satellite signals & Real\tabularnewline

			\bottomrule[1pt] 
			\rowcolor{gray!50}\multicolumn{5}{l}{\textit{Event-triggered \ac{vtr} method}}\tabularnewline
			\hline
			\rowcolor{gray!20} Zhang et al. \cite{zhang2020vehicle} & $>$600 m & Equiped with licence plate
			& Layout traffic cameras on the road with license plate recognition
			functionality & Real\tabularnewline
			
			Michau et al. \cite{michau2017bluetooth}& $>$200 m & Bluetooth & Layout a large number of Bluetooth around the road  & Real\tabularnewline
			
			\rowcolor{gray!20} Wu et al. \cite{wu2014cloud} & $>$250 m & \ac{rfid} & Layout a large number of \ac{rfid} around the road & Simulated\tabularnewline
			
			Gu et al. \cite{gu2021spatio}  & $>$500 m & WiFi & Layout a large number of WiFi around the road & Real\tabularnewline
			
			\bottomrule[1pt] 
			\rowcolor{gray!50}\multicolumn{5}{l}{\textit{Wireless network-assisted \ac{vtr} method}}\tabularnewline
			\hline
			\rowcolor{gray!20} 
			Owen et al. \cite{owen2022vehicle}& 20 m & Cellular device measusing TDoA  & Covered by signals from \acpl{bs} & Simulated \tabularnewline
			
			Rosado et al. \cite{del2016feasibility} & 30 cm & Cellular device measusing TDoA, V2I (using \ac{gps}, \ac{imu}, etc.)
			& Covered by signals from \acpl{bs} & Simulated \tabularnewline
			
			\rowcolor{gray!20} Proposed  & 12.6 m$\sim$14.7 m & Any wireless communication device & Covered by signals from \acpl{bs} & Real\tabularnewline
			
			\bottomrule[1.5pt] 
		\end{tabular}
		\label{tab:comparison}
	\end{table*}

	The \ac{gnss}-aided map-matching method, differing from the \ac{gnss}-aided localization approach, achieves VTR by exploring the temporal relationship between recovered positions along the trajectory using \ac{hmm} techniques. It is the process of aligning a sequence of observed user positions, typically raw GPS data, with the road network on a digital map. Studies like Newson and Krumm \cite{newson2009hidden} apply \ac{hmm} to align GPS data with road networks, considering noise and layout. Lou et al. \cite{lou2009map} introduce ST-Matching for low-sampling-rate GPS trajectories, integrating spatial and temporal analyses to determine the most accurate paths. Hsueh et al. \cite{hsueh2018map} propose to enhance computational efficiency by incorporating directional features and applying GPS clustering \cite{xing2024block,zhao2022analysis,xing2023clustering,zhao2024robust,xing2023blockdiagonal} and smoothing. Jin et al. \cite{jin2022transformer} develop a transformer-based model that uses transfer learning to improve map-matching accuracy with limited labeled data, while Hu et al. \cite{hu2023amm} propose an adaptive online map-matching algorithm that calibrates GPS data dynamically, enhancing the system's accuracy, robustness, and portability through a collaborative evaluation model and self-tuning parameters.

	The event-triggered \ac{vtr} method relies on a network of fixed sensors such as traffic cameras \cite{castillo2008trip, zhaopointlie}, Bluetooth \cite{michau2017bluetooth}, \ac{rfid} \cite{wu2014cloud}, and WiFi \cite{li2022wivelo} deployed across the road network. For instance, the work \cite{castillo2008trip} employs license plate recognition systems in conjunction with traffic cameras to identify vehicles at various key locations and, based on their timestamps, connects these key locations to form a continuous path.
	The work \cite{michau2017bluetooth} tracks vehicles equipped with Bluetooth by deploying numerous Bluetooth scanners around the roads. The vehicle is located near a Bluetooth scanner if the scanner can detect the Bluetooth signal from the vehicle.
	However, event-triggered methods estimate the location of vehicles as the location that is closest to the sensors through sensor-triggered sensing, and then simply connect these triggered positions to form the trajectory of vehicles, resulting in relatively poor \ac{vtr} performance, with reconstruction errors often exceeding 200 meters, as shown in Table~\ref{tab:comparison}.
	Furthermore, event-triggered approaches rely on the widespread deployment of specialized devices like WiFi, Bluetooth, and cameras, which may not always be available in practice. For instance, suburban roads may lack traffic cameras, and most outskirts may not have fixed Bluetooth or WiFi detectors around the roads.

	The widespread deployment of wireless communication networks has facilitated significant advancements in utilizing wireless signal channel characteristics to address real-world challenges \cite{zhao2024huber,dan2024multiple,LiZh:25,zhang2024using,tang2022youcan,li2024deep,xu2024teg}.
	The wireless network-assisted \ac{vtr} method leverages periodic signals from wireless communication devices (e.g., mobile phones, car telephones, mobile routers) on vehicles to determine vehicle trajectories at predefined time intervals. 
	With the widespread application of machine learning and deep learning \cite{zhao2024minimax,
		guo2025bearing,
		LiZh:24,xu2025two,
		li2024optimizing,
		yu2024machine,
		ke2025detection,yu2025identifying,shen2024imagpose, shen2024imagdressing, shenadvancing, shen2024boosting, shen2023pbsl,zhang2022covid , su2022mixed ,shen2024accurate, wang2024deep}, it has become possible to leverage unsupervised learning techniques to address \ac{vtr} technology.
	The works \cite{owen2022vehicle} estimate the location of vehicles at each time slot by analyzing the precise \ac{tdoa} measurement of received signals using a special receiver with mutiple antennas. The vehicle trajectory is given by connecting a series of closely spaced locations, and achieve an error of around 20 meters. In order to improve the  \ac{vtr} performance, the work \cite{del2016feasibility} propose a \ac{v2i}-based method capable of achieving sub-centimeter-level error, but it relies on GPS data. As demonstrated in Table~\ref{tab:comparison}, existing network-assisted methods rely on measuring \ac{tdoa} and, as a result, necessitate specialized decoding receivers capable of deciphering communication protocols. This requirement leads to high application costs. Furthermore, these methods have only been validated through simulation experiments in existing research, rendering them less practical for real-world applications.

	The \ac{rss} is the linear average over the power contributions of the resource elements carrying cell-specific reference signals within the measured frequency bandwidth, which can be measured and reported by any wireless communication device. However, due to its noisy and sporadic nature, it has not been previously utilized for \ac{vtr}.
	We aim to address the VTR problem using traditional machine learning techniques \cite{dan2024evaluation,
		xu2019energy,zhu2025dmaf,
		tang2022few,liu2023spts,li2024research,dan2024image}.
	{Traditional RSS-based localization methods estimate the location at each time slot by weighting BS positions according to RSS, but the resulting locations are often highly scattered and deviate significantly from the  trajectory due to substantial RSS fluctuations. Observed that a) RSS adheres to a consistent path loss model along the trajectory, b) the trajectory is continuous, and c) the vehicle moves along the road network.}
	Existing research has demonstrated that sequential data typically possess sequential characteristics that can be leveraged \cite{xing2024unsupervised,	tang2022optimal,zhao2024multi,
		tang2023character,xu2024spot}.
	This paper seeks to identify a continuous trajectory within a road network that maximizes the probability of collecting sequential \ac{rss} measurements.
	The \ac{rss} is collected sequentially and periodically, which corresponds to a scenario where a vehicle travels on the roads, and the vehicle measures and reports the \ac{rss} of the wireless signal emitted by the neighboring \acpl{bs} at predefined time intervals. 
	Existing Map-matching aligns user positions from raw GPS data with digital map road networks \cite{lou2009map}. In contrast, our method reconstructs vehicle trajectories using \ac{rss} measurements with the constraint that the trajectories lie on an unknown road network, rather than merely aligning positions with existing roads. Therefore, traditional map-matching methods are not applicable to the problem addressed in this paper.
	Consequently, the method needs to confront two significant challenges:
	
	\begin{enumerate}
		\item[$\bullet$] \textit{How to extract the spatio-temporal correlations embedded in the noisy and sporadic \ac{rss} measurement?}
		Wireless signals experience significant interference from electromagnetic sources, leading to substantial noise. 
		Furthermore, signals from a specific \ac{bs} can reliably be received only within its proximity, and even within this proximity, signal reception may be hindered due to irregular obstacles, contributing to a sporadic characteristic in the wireless signal. Consequently, the \ac{rss} displays both noise and sporadic features. Extracting spatio-temporal correlations from sequential \ac{rss} measurements poses a considerable challenge.

		\item[$\bullet$] \textit{How to search for a continuous path within the highly intricate road network?}
		Urban maps often consist of numerous roads. Thus, \ac{vtr} method usually potentially mislocate a moving vehicle outside of the actual road. Integrating map information into the pathfinding process and finding the feasible path efficiently from thousands of roads presents a significant challenge.
	\end{enumerate}

	This paper introduces an innovative \ac{rss}-based \ac{vtr} strategy that leverages the spatio-temporal correlations within \ac{rss} data to generate a continuous road trajectory. 
	The framework comprises two key components: (1) a model-based \ac{rss} embedding method employing \ac{hmm}, and (2) a model-free rough trajectory estimation method constrained by maximum speed.
	Firstly, we propose an integrated \ac{rss} embedding and \ac{vtr} method that alternately learns the \ac{rss} propagation model, mobility model, and reconstructs the trajectory using HMM. 
	The HMM is typically employed for noisy and sparse location data in map-matching problems. In this paper, the HMM is employed for noisy and sporadic \ac{rss} data.
	Specifically, 
	Obstacles, environmental conditions, and electromagnetic interference, along with path loss, result in noisy and sporadic \ac{rss} data for vehicles. The incorporation of the HMM effectively leverages the spatio-temporal correlation of \ac{rss}, while the application of \ac{rss} embedding adeptly addresses the challenges posed by sporadic measurements.
	To ensure the generated trajectory forms a continuous path on the road, we propose a road graph model and incorporate it as a constraint in the \ac{vtr} optimization, ensuring alignment with a feasible path.
	Secondly, we introduce an efficient maximum speed-constrained rough trajectory estimation method, providing the guidance for our alternating optimization model. The maximum speed constraint ensures that \ac{rss} measurements at adjacent times are relatively close, guaranteeing spatial continuity in the generated trajectory.
	
	In conclusion, we make the following contributions:
	
	\begin{enumerate}

		\item[$\bullet$] {\textit{Innovative \ac{rss}-based VTR Scheme}: 
			
			This paper presents the first effort to exclusively utilize \ac{rss} data for \ac{vtr}, pioneering the development of an HRE algorithm that searches for vehicle trajectories along roads based on \ac{rss} measurements. First, we consider optimize signal propagation model parameters and the trajectory with a fixed mobility model.
			Second, we address scenarios where vehicle speed varies without prior information by designing an adaptive mobility model and proposing an optimization method capable of simultaneously optimizing signal propagation model parameters, mobility model parameters, and the trajectory. Third, we have developed a maximum speed-constrained rough trajectory estimation method to guide the HRE algorithm. The proposed HRE method leverages the hidden spatial-temporal correlations within sequential, noisy, and sporadic \ac{rss} data, generating continuous and accurate trajectories.
			
		}
		
		\item[$\bullet$] {\textit{Outstanding VTR Performance}: 
			We generate two datasets by driving in Chengdu and Shenzhen, China, using a smartphone as the receiver and tens of thousands of 5G NR \acpl{bs} as transmitters. These efforts resulted in capturing a 161 km-long trajectory in Chengdu and a 142 km-long trajectory in Shenzhen. Our experiments show that the proposed method demonstrates superior effectiveness on these two datasets compared to state-of-the-art methods and exhibits robust performance even under sporadic \ac{rss} measurements.
		}
		
	\end{enumerate}
	
	The remainder of the paper is structured as follows: Section II presents the system model, including the measurement model of \ac{rss}, \ac{rss} propagation model, and vehicle mobility model. Thereafter, section III presents the design of HMM-based \ac{rss} embedding for \ac{vtr}, as well as a rough trajectory estimation method. In section IV,  we develop performance comparison, parameter sensitivity analysis, and ablation study based on two real datasets.
	Finally, the paper is concluded in section V.

	\section{System model}
	
	\label{sec:System-Model}
	
	\subsection{Measurement Model of \ac{rss} from Each \ac{bs}}

	Consider a wireless communication network with a lot of \acpl{bs}, where \ac{bs}
	serves a mobile vehicle.
	Each \ac{bs} has $N_{\text{t}}$ antennas, and the mobile vehicle
	has $N_{\text{r}}$ antennas. The downlink channel from the $q$th
	\ac{bs} to the mobile vehicle is given by $\mathbf{H}_{q}\in\mathbb{C}^{N_{\text{r}}\times N_{\text{t}}}$.
	Denote $\mathbf{Z}\in\mathbb{C}^{N_{\text{t}}\times N_{\text{t}}}$
	as the \ac{dft} matrix with its $i$th column $\mathbf{z}_{i}$ being
	the $i$th {\em \ac{csi} beam}.
	The \ac{rss} measured at the $j$th receive antenna of the vehicle for the $i$th {\em \ac{csi} beam} transmitted by the $q$th \ac{bs} is defined as the average received signal power
	$\bar{g}_{q,i}^{[j]}=\mathbb{E}\{||\mathbf{e}_{j}^{\text{T}}\mathbf{H}_{q}\mathbf{z}_{i}||^{2}\}$,
	where the superscript $\mathrm{T}$ denotes the transpose, $\mathbf{e}_{j}$
	is a vector of zeros except for the $j$th entry being $1$, and the
	expectation $\mathbb{E}\{\cdot\}$ is taken over the small-scale fading.

	At each time slot, there exist $N_{\text{r}}\times N_{\text{t}}$ \ac{rss} values between the vehicle and a \ac{bs}. However, practical limitations dictate that only the strongest \ac{rss} is reported by the wireless device on the vehicle at each time slot. Let ${\bar{g}_{q,(i)}^{[j]}}$ represent the ordered \ac{rss} values. The strongest \ac{rss} is recorded as $\bar{g}_{q,(1)}^{\mathrm{max}}$. At each time slot, the vehicle records the \ac{rss} values as $\mathbf{y}=\{\bar{g}_{q,(1)}^{\mathrm{max}}\}_{q\in\mathcal{Q}}$, where $\mathcal{Q}$ contains the indices of \acpl{bs} whose signals can be measured by the vehicle.

	\subsection{Signal Propagation Model}
	
	Consider there are $Q$ \acpl{bs} located at position $\{\mathbf{o}_{q}\}_{q=1}^{Q}$, and a vehicle
	travels along a road at time $t=1,2,...,T$. The \ac{rss} 
	$\mathcal{Y}_{t}=(\mathbf{y}_{1},\mathbf{y}_{2},\dots,\mathbf{y}_{t})$
	are measured by the vehicle on the road locations $\mathcal{X}_{t}=(\mathbf{x}_{1},\mathbf{x}_{2},\dots,\mathbf{x}_{t})$, and the measuring time slot is $\delta$ second, where $\mathbf{y}_{t}=\{y_{t,i}\}_{i\in \mathcal{Q}_{t}}$
	is the measured \ac{rss} at time slot $t$ from the \acpl{bs} indexed by the natural number in $\mathcal{Q}_{t}$ when the vehicle is
	located at the position $\mathbf{x}_{t}\in\mathbb{R}^{2}$.
	
	The \ac{rss} of the signal emitted from the $q$th \ac{bs} at time slot $t$ is modeled as $	y_{t,q}=\beta_{q}+\alpha_{q}\mathrm{log}_{10}\|\mathbf{o}_{q}-\mathbf{x}_{t}\|_{2}+\varepsilon$,
	where $\varepsilon\sim\mathcal{N}(0,\sigma_{q}^{2})$ is the shadowing
	related to $q$th \ac{bs}. Thus, the \textit{observation probability} of
	$y_{t,q}$ is calculated by $p(y_{t,q}|\mathbf{x}_{t})  =\frac{1}{(2\pi)^{1/2}\sigma_{q}}\mathrm{exp}\Big(-\frac{1}{2\sigma_{q}^{2}}
	(y_{t,q}-\beta_{q}-\alpha_{q}\mathrm{log}_{10}\|\mathbf{o}_{q}-\mathbf{x}_{t}\|_{2})^{2}\Big)$
	and the \textit{observation probability}
	of $\mathbf{y}_{t}$ is given by 
	\begin{align}
		p(\mathbf{y}_{t}|\mathbf{x}_{t},\bm{\Theta})=\prod_{q\in\mathcal{Q}_{t}}p(y_{t,q}|\mathbf{x}_{t})\label{eq:prob-observ}
	\end{align}
	where $\bm{\Theta}=\{\alpha_{q},\beta_{q},\sigma_{q}\}_{q=1}^{Q}$. The parameter $\bm{\Theta}$ is unknown and needs to be learned using location-labeled \ac{rss}.

	\subsection{Fixed Vehicle Mobility Model}
	\label{subsec:Vehicle-Trajectory-Model}
	
	The road network consists of a collection of interconnected roads,
	forming intersections that enable the selection of multiple travel
	routes along the roads. This paper represents the road network in
	a given map as a road graph $\mathcal{G}=(\mathcal{V},\mathcal{E})$. 
	The location points
	spaced $\gamma$ meters apart on the road are seen as nodes in $\mathcal{V}$. 
	There exists edge between two nodes $\mathbf{s}_i$ and $\mathbf{s}_j$, i.e., $(\mathbf{s},\mathbf{s}_j)\in\mathcal{E}$, if the number of nodes along the minimum traveling path between position $\mathbf{s}_i$ and position $\mathbf{s}_j$ is less than $K$, where $K=\lceil\frac{v_{\mathrm{max}}\delta}{\gamma}\rceil$ is a natural number. 
	The reason is that, if the vehicle is traveling at its maximum speed, $v_{\mathrm{max}}$, it can cover a maximum distance of $K\gamma$ meters within a time slot of $\delta$ seconds. In practice, we usually set $K$ slightly larger than $\lceil\frac{v_{\mathrm{max}}\delta}{\gamma}\rceil$ to allow for vehicles that slightly exceed the maximum speed.

	Any trajectory can be constructed by choosing a sequence of ordered nodes from $\mathcal{V}$ and the edges connecting them.
	We aim to search for a trajectory $\mathcal{X}_T$ within the graph
	$\mathcal{G}$ only based on a sequential \ac{rss} measurement $\mathcal{Y}_T$. Since the measurement in 
	$\mathcal{Y}_T$ is collected every $\delta$ seconds, for any selected
	nodes $\mathbf{s}_{i},\mathbf{s}_{j}\in\mathcal{V}$ for $\mathbf{x}_{t-1},\mathbf{x}_{t}$,
	the driving speed $v$ m/s of the vehicle traveling from $\mathbf{x}_{t-1}=\mathbf{s}_{i}$
	to $\mathbf{x}_{t}=\mathbf{s}_{j}$ in $\delta$ seconds can be calculated.
	This paper defines the transition probability from $\mathbf{x}_{t-1}=\mathbf{s}_{i}$
	to $\mathbf{x}_{t}=\mathbf{s}_{j}$ as the probability
	of vehicle traveling at the speed $v=d(\mathbf{s}_{i},\mathbf{s}_{j})/\delta$ m/s, where $d(\mathbf{s}_{i},\mathbf{s}_{j})$
	measures the routine distance from $\mathbf{s}_{i}$ to $\mathbf{s}_{j}$,
	and can be calculated by counting the number of nodes in the shortest
	path between $\mathbf{s}_{i}$ and $\mathbf{s}_{j}$ in the graph and mutiply $\gamma$. 
	
	Observed that the majority of vehicles travel around an average driving speed, and the number of vehicles traveling at low or high speeds is relatively small. We adopt the Gaussian distribution to model the probability of vehicle driving speeds, i.e., $v\sim\mathcal{N}(v_{\text{avr}},\sigma_{v}^{2})$, where $v_{\text{avr}}$
	is the average driving speed, and $\sigma_{v}$ is the variance. Denote the maximum
	driving speed as $v_{\mathrm{max}}$ m/s, e.g.,
	40$\sim$80 km/h for urban. Suppose the probability of the vehicle traveling in maximum speed limit $v_{\mathrm{max}}$ is $\eta$, thus, the variance can be represented by $\eta$, i.e., 
	\begin{align*}
		\hat{\sigma}_{v}^{2}(\eta)=-\frac{(v_{max}-v_{\text{avr}})^{2}}{W_{-1}(-2\pi\eta^{2}(v_{max}-v_{\text{avr}})^{2})}
	\end{align*}
	where $W_{-1}(\cdot)$ is Lambert $W$ function.
	Then, the \textit{transition probability} of traveling from $\mathbf{x}_{t-1}=\mathbf{s}_{i}$ to $\mathbf{x}_{t}=\mathbf{s}_{j}$
	can be given by
	\begin{align}
		p(\mathbf{x}_{t}=\mathbf{s}_{j}|&\mathbf{x}_{t-1}=\mathbf{s}_{i})=p\Big(v=\frac{d(\mathbf{s}_{i},\mathbf{s}_{j})}{\delta}\Big) \label{eq:prob-trans}\\
		& =\frac{1}{\sqrt{2\pi\hat{\sigma}_{v}^{2}(\eta)}}\mathrm{exp}\Big(-\frac{1}{2\hat{\sigma}_{v}^{2}(\eta)}(\frac{d(\mathbf{s}_{i},\mathbf{s}_{j})}{\delta}-v_{\text{avr}})^{2}\Big)\nonumber 
	\end{align}
	The probability in \eqref{eq:prob-trans} is determined by three parameters: the maximum speed limit $v_{\mathrm{max}}$, the probability of traveling at maximum speed $\eta$, and the average speed $v_{\text{avr}}$. We assume that these three parameters are given.
	
	We assume that we have prior knowledge of the type of road environment on which the trajectory is running. For highway environments, we usually set \( v_{\text{avr}} = 80 \sim 120 \) km/h and \( v_{\text{max}} = 150 \) km/h; for suburban roads, we set \( v_{\text{avr}} = 40 \sim 80 \) km/h and \( v_{\text{max}} = 100 \) km/h; for dense urban areas, we set \( v_{\text{avr}} = 20 \sim 60 \) km/h and \( v_{\text{max}} = 80 \) km/h.

	\subsection{Adaptive Vehicle Mobility Model}
	\label{subsec:Adaptive-Vehicle-Trajectory-Model}
	This paper also considers scenarios where the mobility model parameters are not provided and need to be estimated. Since the mobility velocity is time-varying and unknown in our problem, we face the challenge of modeling the mobility velocity in segments rather than uniformly. We leverage information about speed variations obtained from the variation rates in the RSS data, group the RSS data accordingly, and apply a separate mobility model to each group.

	Specifically, we first propose to evaluate the variation in RSS over consecutive time intervals while accounting for differences in the set of observable BSs. We define the \emph{normalized signal difference} as follows. Let the sets of BSs whose signals are observed at time $t$ and $t+1$ be denoted by $Q_t$ and $Q_{t+1}$, respectively. 
	
	To normalize this measure and account for differences in the number of observable BSs, we compute the \emph{normalized signal difference} as:
	\[
	\rho_t= \frac{\sum_{q \in Q_t \cap Q_{t+1}} |y_{t+1,q} - y_{t,q}|}{|Q_t \cap Q_{t+1}|}, \quad t \in \{1,2,\ldots,T-1\}
	\]
	where $|Q_t \cap Q_{t+1}|$ denotes the cardinality of the set of common BSs. This normalization is suitable for scenarios where the sets of observed BSs at consecutive time intervals differ significantly due to environmental or system-level factors and ensures that comparisons are made only for signals that are present in both time intervals.
	
	Obviously, if the vehicle speed is relatively high, the positions where RSS is collected at consecutive time instants will be farther apart. Therefore, the collected signals' normalized signal difference will be larger. Consequently, we consider dividing $\{\rho_t\}_{t=1}^{T-1}$ into $A$ groups. Define $\rho_{\text{max}} = \max_t \rho_t$ and $\rho_{\text{min}} = \min_t \rho_t$. The index set of the $a$-th group is defined as
	\begin{align*}
		G_{a}=&\Big\{ t|(a-1)\frac{(\rho_{\text{max}}-\rho_{\text{min}})}{A}+\rho_{\text{min}}\leq\rho_{t}\\
		&<a\frac{(\rho_{\text{max}}-\rho_{\text{min}})}{A}+\rho_{\text{min}},t=1,2,...,T-1\Big\}
	\end{align*}

	Suppose the time slots within each group follow consistent mobility model parameters. Then, the transition probability in \eqref{eq:prob-trans} can be written as
	\begin{align}
		&\tilde{p}(\mathbf{x}_{t} = \mathbf{s}_{j} \mid \mathbf{x}_{t-1} = \mathbf{s}_{i}) \label{eq:transition2}\\
		&= \sum_{a=1}^{A} \mathbb{I}_{\{t-1 \in G_{a}\}} \frac{1}{\sqrt{2\pi\sigma_{v}^{2}}} \exp\left( -\frac{1}{2\sigma_{a,v}^{2}} \left( \frac{d(\mathbf{s}_{i}, \mathbf{s}_{j})}{\delta} - v_{a,\text{avr}} \right)^{2} \right) \nonumber
	\end{align}
	where $\mathbb{I}_{\{t-1 \in G_{a}\}}=1$ if $t-1 \in G_{a}$ is true and zero otherwise. Denote the mobility parameter set as $\bm{\Phi}=\{v_{a,\text{avr}},\sigma_{a,v}^{2}\}_{a=1}^{A}$, which is needed to be estimated.

	\section{Problem Formulation}
	
	\begin{figure}
		\centering{}\includegraphics[width=1\columnwidth]{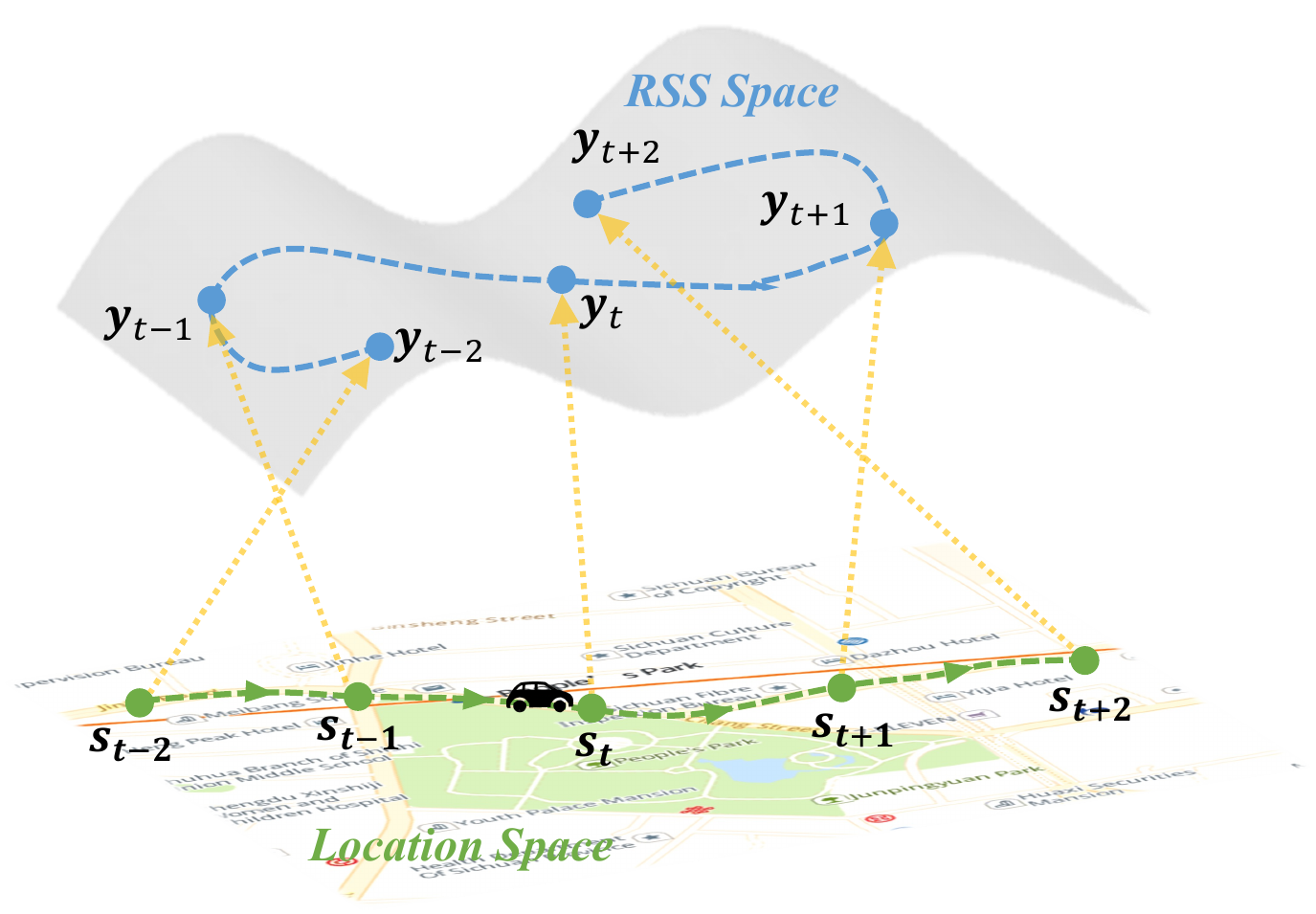}\caption{{Illustration of the relationship between location space and \ac{rss} space: The relationship indicated by yellow arrows is modeled by \ac{rss} propagation, while the relationship depicted by green arrows is modeled by vehicle mobility.}
			\label{fig:Mobility}}
	\end{figure}
	
	As depicted in Figure~\ref{fig:Mobility}, there exists a feasible driving path, which may be quite circuitous, along an unknown road, represented by a series of nodes $\mathbf{s}_{t-2},\mathbf{s}_{t-1},\mathbf{s}_{t},...$. Our goal is to determine the most likely path through this lattice by selecting one position from $\mathcal{V}$ at each time step. This path should strike a balance between responding to the measurements and adhering to the plausibility of the chosen positions. This balance is determined by the probabilities associated with the measurements (red arrows) and the probabilities governing the transitions between the selected positions (yellow arrows) at each time step.
	
	We aim to select a sequence of nodes from $\mathcal{V}$ to create a trajectory $\mathbf{x}_{1}\rightarrow\mathbf{x}_{2}\rightarrow\cdots\rightarrow\mathbf{x}_{T}$ in such a way that it maximizes the probability of collecting the sequential data $\mathbf{y}_{1}\rightarrow\mathbf{y}_{2}\rightarrow\cdots\rightarrow\mathbf{y}_{T}$.
	An HMM-Based \ac{rss} embedding problem with unknown trajectory
	and propagation model parameter is developed.
	Firstly, the probability
	$p(\mathcal{X}_{t},\mathcal{Y}_{t})$ can be written as $ p(\mathcal{X}_{t},\mathcal{Y}_{t})=p(\mathbf{y}_{t}|\mathbf{x}_{t})p\{\mathbf{x}_{t}|\mathcal{Y}_{t-1},\mathcal{X}_{t-1})p(\mathcal{X}_{t-1},\mathcal{Y}_{t-1})$.
	Since the position $\mathbf{x}_{t}$ is only determined by the position
	$\mathbf{x}_{t-1}$ it travels from, the probability $p\{\mathbf{x}_{t}|\mathcal{Y}_{t},\mathcal{X}_{t-1}\}$ can be reduced to $p(\mathbf{x}_{t}|\mathbf{x}_{t-1})$.
	Thus, the probability $p(\mathcal{X}_{t},\mathcal{Y}_{t})$
	can be written as $p(\mathcal{X}_{t},\mathcal{Y}_{t})=p(\mathbf{x}_{t}|\mathbf{x}_{t-1})p(\mathbf{y}_{t}|\mathbf{x}_{t})p(\mathcal{X}_{t-1},\mathcal{Y}_{t-1})$.
	Obviously, there exists a recurrence relation between $p(\mathcal{X}_{t},\mathcal{Y}_{t})$
	and $p(\mathcal{X}_{t-1},\mathcal{Y}_{t-1})$, and
	the solution of maximizing the log-joint probability $\text{log}\:p(\mathcal{X}_{T},\mathcal{Y}_{T})$ is given by the following HMM-based \ac{rss} embedding problem:
	\begin{align}
		\underset{\mathcal{X}_{T},\bm{\Theta}}{\mathrm{maximize}} & \quad\text{log}\Big\{\prod_{t=1}^{T}p(\mathbf{y}_{t}|\mathbf{x}_{t},\bm{\Theta})\prod_{t=2}^{T}p(\mathbf{x}_{t}|\mathbf{x}_{t-1})p(\mathbf{x}_{1})\Big\}\label{eq:prob-J}\\
		\mathrm{subject\:to} & \;\;\mathbf{x}_{t}\in\mathcal{\mathcal{V}},\quad t=1,2,\dots,T\nonumber\\ 
		&\;\;(\mathbf{x}_{t},\mathbf{x}_{t-1})\in\mathcal{E},\quad t=2,\dots,T.\nonumber
	\end{align}
	where the initial probability $p(\mathbf{x}_{1})$ is set to be uniform because there is not prior knowledge of the trajectory in our problem.
	The observation probability $p(\mathbf{y}_{t}|\mathbf{x}_{t},\bm{\Theta})$ in \eqref{eq:prob-J} is given by \eqref{eq:prob-observ}, and the transition probability $p(\mathbf{x}_{t}|\mathbf{x}_{t-1})$ in \eqref{eq:prob-J} is given by \eqref{eq:prob-trans}.
	
	Consider the HMM-based RSS embedding with adaptive vehicle mobility model problem:
	\begin{align}
		\underset{\mathcal{X}_{T},\bm{\Theta},\bm{\Phi}}{\mathrm{maximize}} & \quad\text{log}\Big\{\prod_{t=1}^{T}p(\mathbf{y}_{t}|\mathbf{x}_{t},\bm{\Theta})\prod_{t=2}^{T}\tilde{p}(\mathbf{x}_{t}|\mathbf{x}_{t-1})p(\mathbf{x}_{1})\Big\}\label{eq:prob-J2}\\
		\mathrm{subject\:to} & \;\;\mathbf{x}_{t}\in\mathcal{\mathcal{V}},\quad t=1,2,\dots,T\nonumber\\ 
		&\;\;(\mathbf{x}_{t},\mathbf{x}_{t-1})\in\mathcal{E},\quad t=2,\dots,T.\nonumber
	\end{align}
	where three sets of parameters need to be determined, including the trajectory $\mathcal{X}_{T}$, the signal propagation parameters $\bm{\Theta}$, and the mobility model parameters $\bm{\Phi}$.

	\section{Methodology}
	
	\label{subsec:Alternative-Algorithm}

	\subsection{General Framework}

	{In a multi-variable optimization problem, the alternating optimization strategy works by iteratively optimizing one set of variables while keeping others fixed, thus progressively approaching a global solution. Its advantage lies in breaking down a complex problem into simpler subproblems. In alternating optimization, the original problem is divided into smaller subproblems, each involving a subset of the variables. The process begins with an initial guess for one subset of variables.}

	Observed that there are two blocks of variables in (\ref{eq:prob-J}).
	The variable $\mathcal{X}_T$ is dependent on $\bm{\Theta}$, and $\bm{\Theta}$
	is dependent on $\mathcal{X}_T$. Thus, we propose an alternative optimization method to solve problem (\ref{eq:prob-J}). Specifically,
	problem (\ref{eq:prob-J}) can be divided into two subproblems:
	\begin{align}
		\underset{\{\alpha_{q},\beta_{q},\sigma_{q}\}_{q=1}^{J}}{\mathrm{maximize}} & \quad\text{log}\prod_{t=1}^{T}\prod_{q\in\mathcal{Q}_{t}}p(y_{t,q}|\mathbf{x}_{t},\alpha_{q},\beta_{q},\sigma_{q})\label{eq:prob-J-1}
	\end{align}
	and
	\begin{align}
		\underset{\mathcal{X}_{T}}{\mathrm{maximize}} &
		\quad\sum_{t=1}^{T}\text{log}\:p(\mathbf{y}_{t}|\mathbf{x}_{t},\bm{\Theta})+\sum_{t=2}^{T}\text{log}\:p(\mathbf{x}_{t}|\mathbf{x}_{t-1})
		\label{eq:prob-J-2}\\
		\mathrm{subject\:to} & \quad\mathbf{x}_{t}\in\mathcal{V},\quad t=1,2,....,T\nonumber\\
		&\;\;(\mathbf{x}_{t},\mathbf{x}_{t-1})\in\mathcal{E},\quad t=2,\dots,T.\nonumber
	\end{align}

	We start from an initial trajectory, and solve problems (\ref{eq:prob-J-1}) and (\ref{eq:prob-J-2}) alternatively until the trajectory can not be updated. This alternating optimization has some admirable advantages: firstly,
	problem \eqref{eq:prob-J-1} has a closed-form solution; secondly,
	while (\ref{eq:prob-J-2}) is non-convex, it can be solved based on
	\ac{hmm} decoding algorithm with a globally optimal guarantee.

	\begin{algorithm}
		\textit{Input:} $\mathcal{Y}_T$%, tolerance $\epsilon$
		
		Initialize $\hat{\mathcal{X}}^{(0)}$ with a rough trajectory using Algorithm \ref{alg:initial}. Initialize $v_{\mathrm{max}},\eta,v_{\text{avr}}$ with prior knowledge of the trajectory. 
		
		Repeat
		\begin{itemize}
			\item Solve (\ref{eq:prob-J-1}) \ac{wrt} $\hat{\bm{\Theta}}^{(m+1)}$ with given $\hat{\mathcal{X}}^{(m)}$ based on (\ref{eq:Theta-solution}).
			\item Solve (\ref{eq:prob-J-2}) \ac{wrt} $\hat{\mathcal{X}}^{(m+1)}$ with given $\hat{\bm{\Theta}}^{(m+1)}$ using Algorithm \ref{alg:viterbi}.
		\end{itemize}
		Until $\sum_{t=1}^{T}\|\mathbf{x}_{t}^{(m+1)}-\mathbf{x}_{t}^{(m)}\|_{2}=0$
		
		\textit{Output:} the trajectory $\hat{\mathcal{X}}^{(m+1)}$
		
		\caption{HMM-based \ac{rss} embedding (HRE) algorithm.\label{alg:overal}}
	\end{algorithm}

	The overall algorithm is shown in Algorithm \ref{alg:overal}. The alternating optimization algorithm ensures convergence because the solutions to problem \eqref{eq:prob-J-1} and problem \eqref{eq:prob-J-2} are both optimal. However, it's worth noting that problem \eqref{eq:prob-J} is non-convex, and our alternating algorithm will converge to a suboptimal solution with a bad initial trajectory. Thus, we initially create a rough trajectory to guide the alternating optimization procedure. The initial trajectory $\hat{\mathcal{X}}^{(0)}$ will be provided by Algorithm~\ref{alg:initial} in Section \ref{subsec:Location-Initialization}.

	\subsection{Learning Parameters of the \ac{rss} Propagation Model with Known Trajectory}
	\label{sec:propogation-learn}
	We choose the \ac{rss} which is measured by the $q$th \ac{bs}, i.e.,  $(\tilde{y}_{q,1},\tilde{y}_{q,2},...,\tilde{y}_{q,N_{q}})$ with corresponding measuring position $(\tilde{\mathbf{x}}_{q,1},\tilde{\mathbf{x}}_{q,2},...,\tilde{\mathbf{x}}_{q,N_{q}})$, where $N_{q}$ is the number of \ac{rss} measured by the $q$th \ac{bs}.
	
	Problem (\ref{eq:prob-J-1}) is equivalent to maximize
	\begin{align}
		& \mathrm{log}\prod_{t=1}^{T}\prod_{q\in\mathcal{Q}_{t}}p(y_{t,q}|\mathbf{x}_{t},\alpha_{q},\beta_{q},\sigma_{q})\label{eq:J}\\
		&=\mathrm{log}\prod_{q=1}^{Q}\prod_{i=1}^{N_{q}}p(\tilde{y}_{q,i}|\tilde{\mathbf{x}}_{q,i},\alpha_{q},\beta_{q},\sigma_{q})\nonumber\\
		& =-\sum_{q=1}^{Q}\Big[\sum_{t=1}^{N_q}\Big(\frac{1}{2}\mathrm{ln}2\pi+\mathrm{ln}\sigma_{q}\Big)-\frac{1}{2\sigma_{q}^{2}}\big\|\mathbf{C}_{q}\Big[\begin{array}{c}
			\alpha_{q}\\
			\beta_{q}
		\end{array}\Big]-\mathbf{h}_{q}\big\|_{2}^{2}\Big]\nonumber 
	\end{align}
	where
	\begin{align}
		\label{eq:alp_beta}% zheng remove label
		\mathbf{C}_{q}=\left[\begin{array}{cc}
			\mathrm{log}_{10}\|\mathbf{o}_{q}-\mathbf{\tilde{x}}_{q,1}\|_{2} & 1\\
			\mathrm{log}_{10}\|\mathbf{o}_{q}-\mathbf{\tilde{x}}_{q,2}\|_{2} & 1\\
			\vdots & \vdots\\
			\mathrm{log}_{10}\|\mathbf{o}_{q}-\mathbf{\tilde{x}}_{q,N_q}\|_{2} & 1
		\end{array}\right],\mathbf{h}_{q}=\left[\begin{array}{c}
			{\tilde{y}}_{q,1}\\
			{\tilde{y}}_{q,2}\\
			\vdots\\
			{\tilde{y}}_{q,N_q}
		\end{array}\right].
	\end{align}
	The solution to equation (\ref{eq:prob-J-1}) is obtained by taking
	the derivative of (\ref{eq:J}) \ac{wrt} $[\alpha_{q},\beta_{q}]^{\text{T}}$, $\sigma_{q}^{2}$ and setting it equal to zero, which
	leads to the solution: 
	
	\begin{equation}
		\Big[\begin{array}{c}
			\hat{\alpha}_{q}\\
			\hat{\beta}_{q}
		\end{array}\Big]=(\mathbf{C}_{q}^{\mathrm{T}}\mathbf{C}_{q})^{-1}\mathbf{C}_{q}^{\mathrm{T}}\mathbf{h}_{q},\:\hat{\sigma}_{q}^{2}=\frac{1}{T}\big\|\mathbf{C}_{q}\big[\begin{array}{c}
			\hat{\alpha}_{q}\\
			\hat{\beta}_{q}
		\end{array}\big]-\mathbf{h}_{q}\big\|_{2}^{2}\label{eq:Theta-solution}
	\end{equation}
	with linear computational complexity $\mathcal{O}(QT)$.
	\subsection{Reconstructing the Trajectory with Known \ac{rss} Propagation Model }
	\label{sec:traj-learn}
	
	\begin{algorithm}
		\textit{Input:} the \ac{rss} measurement $\mathcal{Y}_{T}$, the \ac{rss} propagation model $\bm{\Theta}$, and the equal spacing $\gamma^{(1)}$, $\gamma^{(2)}$

		{For $m=1,2$}
		\begin{itemize}
			\item Construct the graph $\ensuremath{\mathcal{G}^{(m)}=(\mathcal{V}^{(m)},\mathcal{E}^{(m)})}$,
			where $\mathcal{V}^{(m)}$ is constructed based on $\gamma^{(m)}$,
			and $\mathcal{E}^{(m)}$ is constructed based on $K^{(m)}=\lceil\frac{v_{\mathrm{max}}\delta}{\gamma^{(m)}}\rceil$.
			
			\item Initialize the path set $P(j)=\{j\}$,
			and the objective function loss $L(1,j)=\text{log}\:p(\mathbf{y}_{1}|\mathbf{s}_{j},\bm{\Theta})$ in the state $\mathbf{s}_{j}$ at time slot $1$, $j=1,2,\dots,N_{\text{node}}^{(m)}$, where $N_{\text{node}}^{(m)}$ is the number of nodes in the $\mathcal{V}^{(m)}$.
			
			\item For each $i=2,3,....,T$
			\begin{itemize}
				\item For each $j=1,2,\dots,N_{\text{node}}^{(m)}$
				\begin{enumerate}
					\item For each $k=1,2,\dots,N_{\text{node}}^{(m)}$
					
					If $(\mathbf{s}_{k},\mathbf{s}_{j})\in\mathcal{E}^{(m)}$: $l(k)=L(i-1,k)+\text{log}\:p(\mathbf{y}_{i}|\mathbf{s}_{j},\bm{\Theta})+\text{log}\:p(\mathbf{s}_{j}|\mathbf{s}_{k})$
					\item Choose the optimal path if terminating in the state $\mathbf{s}_{j}$ at time slot $i$: $\hat{k}=\text{argmin}_{k\in\{1,2,...,N_{\text{node}}^{(m)}\}}l(k)$
					\item Calculate the loss $L(i,j)=l(\hat{k})$
					\item Update the path set $P(j)=P(\hat{k})[1:i-1]\cup \{j\}$
				\end{enumerate}
			\end{itemize}
			\item Path $\mathcal{X}_{T}^{(m)}=P(\text{argmin}_{j\in\{1,2,...,N_{\text{node}}^{(m)}\}}L(T,j)$
			
		\end{itemize}

		\textit{Output:} $\mathcal{X}_{T}^{(m)}$
		
		\caption{Two-stage forward-backward algorithm for \ac{vtr} with known \ac{rss} propagation model.\label{alg:viterbi}}
	\end{algorithm}

	Problem (\ref{eq:prob-J-2}) can be solved using the recursive and backtracking strategy in Algorithm \ref{alg:viterbi}. Specifically, Algorithm \ref{alg:viterbi} begins by recursively computing the maximum possible score for each location state at every time step. This is achieved by considering the score obtained at the previous time step, along with the state transition probabilities and emission probabilities. For each time step and each state, our algorithm calculates a score, given by $\text{log}\:p(\mathbf{y}_{t}|\mathbf{x}_{t},\bm{\Theta})+\text{log}\:p(\mathbf{x}_{t}|\mathbf{x}_{t-1})$ (note that, at time slot $t=1$, the score is solely $\text{log}\:p(\mathbf{y}_{t}|\mathbf{x}_{t},\bm{\Theta})$). This score represents the probability of achieving the best path to that state given the observed sequence. The score is computed by considering the previous time step's state scores, adding the state transition probability and emission probability. The state with the highest score is then selected as the best state.
	After completing the recursive process, our algorithm performs backtracking. It starts at the last time step, selecting the state with the highest score, and then traces back to the previous time step, repeating the process of choosing the best state. This continues until it reaches the first time step, thereby determining the most probable state sequence.

	In Algorithm \ref{alg:viterbi}, any feasible route consists of $T$
	moves. We start from one of $N_{\text{node}}^{(m)}$ nodes in the graph, and at each following position, there are at most $K$ candidate
	positions for the next move. Thus, the overal time complexity is $\mathcal{O}(N_{\text{node}}^{(m)}KT)$. Observed that a very small  $\gamma^{(m)}$ and an extremely large map can result in an exceptionally large $N_{\text{node}}^{(m)}$, thereby leading to a significantly large computational complexity.
	To reduce complexity, our strategy involves initially setting a relatively large $\gamma^{(1)}$ value to obtain a coarse path. This larger $\gamma^{(1)}$ value results in a smaller value for $N$. After obtaining a coarse path, we then utilize this coarse path along with a smaller $\gamma^{(2)}$ value to create a narrower, more refined graph. Consequently, the overall algorithm complexity is $\mathcal{O}((N_{\text{node}}^{(1)}+N_{\text{node}}^{(2)})KT)$. The time complexity is linear with the sample number $T$ if $\gamma^{(1)}$ and $\gamma^{(2)}$ are appropriately chosen.

	\subsection{Convergence Analysis}
	The convergence of the proposed alternating optimization algorithm is ensured by iteratively solving two interdependent subproblems in (\ref{eq:prob-J}). Specifically, problem (\ref{eq:prob-J-1}) optimizes the signal propagation parameters \(\bm{\Theta}\) with a closed-form solution, while problem (\ref{eq:prob-J-2}) optimizes the trajectory \(\mathcal{X}_T\) using an \ac{hmm}-based decoding algorithm with a globally optimal guarantee. Consequently, solving problems (\ref{eq:prob-J-1}) and (\ref{eq:prob-J-2}) consistently increases the objective function value of (\ref{eq:prob-J}). Although the overall problem (\ref{eq:prob-J}) is non-convex, the algorithm progressively refines the solutions to these subproblems, ensuring that the objective function value does not decrease in each iteration.

	\subsection{HRE with Adaptive Mobility (HREA)}
	
	Consider solving problem (\ref{eq:prob-J2}). Similar to problem (\ref{eq:prob-J}), problem (\ref{eq:prob-J2}) involves three blocks of variables. The variable $\mathcal{X}_T$ depends on $\bm{\Theta}$ and $\bm{\Phi}$, and vice versa. The variables $\bm{\Theta}$ and $\bm{\Phi}$ are independent of each other. Therefore, we propose an alternative optimization method to solve problem (\ref{eq:prob-J2}). Specifically, problem (\ref{eq:prob-J2}) can be divided into three subproblems:
	\begin{align}
		\underset{\{\alpha_{q},\beta_{q},\sigma_{q}\}_{q=1}^{J}}{\mathrm{maximize}} & \quad\text{log}\prod_{t=1}^{T}\prod_{q\in\mathcal{Q}_{t}}p(y_{t,q}|\mathbf{x}_{t},\alpha_{q},\beta_{q},\sigma_{q})\label{eq:prob-J1-1}
	\end{align}
	\begin{align}
		\underset{\{v_{a,\text{avr}},\sigma_{a,v}^{2}\}_{a=1}^{A}}{\mathrm{maximize}} & \quad\text{log}\prod_{t=1}^{T}\tilde{p}(\mathbf{x}_{t}|\mathbf{x}_{t-1})\label{eq:prob-J1-2}
	\end{align}
	and
	\begin{align}
		\underset{\mathcal{X}_{T}}{\mathrm{maximize}} &
		\quad\sum_{t=1}^{T}\text{log}\:p(\mathbf{y}_{t}|\mathbf{x}_{t},\bm{\Theta})+\sum_{t=2}^{T}\text{log}\:\tilde{p}(\mathbf{x}_{t}|\mathbf{x}_{t-1})
		\label{eq:prob-J1-3}\\
		\mathrm{subject\:to} & \quad\mathbf{x}_{t}\in\mathcal{V},\quad t=1,2,....,T\nonumber\\
		&\;\;(\mathbf{x}_{t},\mathbf{x}_{t-1})\in\mathcal{E},\quad t=2,\dots,T.\nonumber
	\end{align}

	We can solve problem (\ref{eq:prob-J1-1}) using method in Sec. \ref{sec:propogation-learn}, and solve problem (\ref{eq:prob-J1-3}) using method in Sec. \ref{sec:traj-learn}. For solving problem (\ref{eq:prob-J1-2}), we consider the following equivalent problem:
	\begin{align*}
		\underset{\{v_{a,\text{avr}},\sigma_{a,v}^{2}\}_{a=1}^{A}}{\mathrm{maximize}} & \quad-\sum_{t=2}^{T}\sum_{a=1}^{A}\mathbb{I}_{\{t-1\in G_{a}\}}\frac{1}{2}\log(2\pi\sigma_{v}^{2})\nonumber\\
		&+\frac{1}{2\sigma_{a,v}^{2}}\left(\frac{d(\mathbf{x}_{t-1},\mathbf{x}_{t})}{\delta}-v_{a,\text{avr}}\right)^{2}
	\end{align*}
	which can be written as
	\begin{align*}
		\underset{v_{a,\text{avr}},\sigma_{a,v}^{2}}{\mathrm{minimize}}\:\sum_{t\in G_{a}}\frac{1}{2}\log(2\pi\sigma_{v}^{2})+\frac{1}{2\sigma_{a,v}^{2}}\left(\frac{d(\mathbf{x}_{t},\mathbf{x}_{t+1})}{\delta}-v_{a,\text{avr}}\right)^{2}
	\end{align*}
	for $a=1,2,...,A$. The solution can be given by the first-order conditions:
	\begin{align}
		v_{a,\text{avr}} = \frac{1}{|G_{a}|} \sum_{t \in G_{a}} \frac{d(\mathbf{x}_{t}, \mathbf{x}_{t+1})}{\delta}\label{eq:velocity}
	\end{align}
	\begin{align}
		\sigma_{a,v}^{2} = \frac{1}{|G_{a}|} \sum_{t \in G_{a}} \left( \frac{d(\mathbf{x}_{t}, \mathbf{x}_{t+1})}{\delta} - v_{a,\text{avr}} \right)^{2}.\label{eq:sigma}
	\end{align}

	The entire procedure is shown in Algorithm \ref{alg:overal2}. Problems (\ref{eq:prob-J1-1}) and (\ref{eq:prob-J1-2}) optimize the signal propagation parameters \(\bm{\Theta}\) and the mobility parameter \(\bm{\Phi}\) with closed-form solutions, while problem (\ref{eq:prob-J1-3}) optimizes the trajectory \(\mathcal{X}_T\) using an \ac{hmm}-based decoding algorithm that guarantees global optimality. Consequently, solving problems (\ref{eq:prob-J1-1}), (\ref{eq:prob-J1-2}), and (\ref{eq:prob-J1-3}) consistently increases the objective function value of (\ref{eq:prob-J2}). Thus, Algorithm \ref{alg:overal2} is convergent.

	\begin{algorithm}
		\textit{Input:} $\mathcal{Y}_T$%, tolerance $\epsilon$
		
		Initialize $\hat{\mathcal{X}}^{(0)}$ with a rough trajectory using Algorithm \ref{alg:initial}.
		Repeat
		\begin{itemize}
			\item Solve (\ref{eq:prob-J1-1}) \ac{wrt} $\hat{\bm{\Theta}}^{(m+1)}$ with given $\hat{\mathcal{X}}^{(m)}$ based on (\ref{eq:Theta-solution}).
			\item Solve (\ref{eq:prob-J1-2}) \ac{wrt} $\hat{\bm{\Phi}}^{(m+1)}$ with given $\hat{\mathcal{X}}^{(m)}$ based on (\ref{eq:velocity}) and (\ref{eq:sigma}).
			\item Solve (\ref{eq:prob-J1-3}) \ac{wrt} $\hat{\mathcal{X}}^{(m+1)}$ with given $\hat{\bm{\Theta}}^{(m+1)}$ and $\hat{\bm{\Phi}}^{(m+1)}$ using Algorithm \ref{alg:viterbi}.
		\end{itemize}
		Until $\sum_{t=1}^{T}\|\mathbf{x}_{t}^{(m+1)}-\mathbf{x}_{t}^{(m)}\|_{2}=0$%\leq\epsilon$
		
		\textit{Output:} the trajectory $\hat{\mathcal{X}}^{(m+1)}$
		\caption{HREA algorithm.\label{alg:overal2}}
	\end{algorithm}

	\subsection{Fast Continuous Trajectory Reconstructing}
	\label{subsec:Location-Initialization}
	
	{The performance of alternating optimization is highly dependent on the initial guess. Therefore, the objective of this section is to provide an effective initial setting for the proposed alternating optimization algorithm.}
	
	To efficiently establish a preliminary trajectory based solely on \ac{rss} measurements, without utilizing a road graph, we estimate the vehicle's position at each time slot by weighting the surrounding \acpl{bs} positions. A rough trajectory is then generated by applying a speed constraint to these estimated positions.
	
	Specifically, 
	we estimate the position of the vehicle as the positions of the nearest \ac{bs}, i.e., $\mathbf{z}_{t}=\mathbf{o}_{q^{*}}$, where $q^{*}={\mathrm{argmax}}_{q\in\mathcal{Q}}\:y_{t,q}$.
	The estimated positions $\{\mathbf{z}_{t}\}$ is merely coarse approximations
	of the real position of vehicles. Observed that the estimated
	positions $\{\mathbf{z}_{t}\}$ are highly discrete and do not appear
	to resemble a coherent trajectory on the road. 
	During the process of estimating positions based on \ac{rss}
	measurements, a scenario might arise where the estimated position
	at time $t$ lies on road A, while the subsequent estimated position
	at $t+1$ unexpectedly jumps to a distant road B. Given the constraints
	imposed by the vehicle's travel speed, such sudden transitions are
	implausible. Thus, the challenge lies in ensuring that temporally
	adjacent \ac{rss} measurements yield spatially adjacent estimated positions.
	We aim to derive a continuous trajectory $\mathcal{X}_{T}$ in geographical
	space based on estimated positions $\{\mathbf{z}_{t}\}$. The maximum speed-constrained 
	rough trajectory estimation problem is formulated as minimizing $\sum_{t=1}^{T}\|\mathbf{x}_{t}-\mathbf{z}_{t}\|_{2}$ with the constraint that $\|\mathbf{x}_{t}-\mathbf{x}_{t-1}\|_{2}\leq v_{\mathrm{max}}\delta,\forall t\in\{2,3,...,T\}$, where $v_{\mathrm{max}}\delta$ is the maximum traveling distance between adjacent positions
	in $\delta$ seconds.

	\begin{algorithm}
		\textit{Input:}  $\{\mathbf{z}_t\}_{t=1}^{T}$, and $v_{\text{max}\delta}$
		
		Initialize a strictly feasible trajctory $\mathcal{X}^{(0)}$, $c^{(0)}>0$,
		$\mu>1$, tolerance $\epsilon_{1},\epsilon_{2}>0$ according to \cite{boyd2004convex}.	
		Repeat
		\begin{itemize}
			\item Centering step:
			Initialize $\mathcal{X}^{(m,0)}=\mathcal{X}^{(m)}$.
			\begin{itemize}
				\item Repeat
				\begin{itemize}
					\item $p^{(n)}=\mathbf{H}^{-1}(\mathcal{X}^{(m,n)},c^{(m)})\varphi'(\mathcal{X}^{(m,n)},c^{(m)})$
					\item $\mathcal{X}^{(m,n+1)}=\mathcal{X}^{(m,n)}-p^{(n)}$
				\end{itemize}
				\item Until $\|\varphi'(\mathcal{X}^{(m,n)},c^{(m)})\|_2<\epsilon_{1}$
				\item Update $\mathcal{X}^{(m+1)}=\mathcal{X}^{(m,n)}$
			\end{itemize}
			\item Increase $c^{(m+1)}:=\mu c^{(m)}$
		\end{itemize}
		Until $c^{(m)}\geq(T-1)/\epsilon_{2}$
		
		\textit{Output:} $\mathcal{X}^{(m+1)}$

		\caption{Maximum speed-constrained recovery (MSR).\label{alg:initial}}
		
	\end{algorithm}
	The proposed problem contains a quadratic objective function and a quadratic inequality constraint. Therefore, the problem is convex and can be solved using conventional convex optimization methods, such as the interior point method.  Specifically, the problem can be written as
	\begin{align}
		\underset{\mathcal{X}_{T}}{\mathrm{minimize}} & \quad\varphi(\mathcal{X}_{T},c)\label{eq:P_qp1-2}
	\end{align}
	where $c$ is a penalty coefficient, and
	\[
	\varphi(\mathcal{X}_{T},c)=\sum_{t=1}^{T}c\|\mathbf{x}_{t}-\mathbf{z}_{t}\|_{2}-\sum_{t=2}^{T}\mathrm{log}\:\left[-\|\mathbf{x}_{t}-\mathbf{x}_{t-1}\|_{2}+v_{\mathrm{max}}\delta\right]
	\]
	The gradient of $\varphi(\mathcal{X}_{T},c)$ \ac{wrt} $\mathcal{X}_{T}$ is given by $\varphi'(\mathcal{X}_{T},c)=[\frac{\partial\varphi(\mathcal{X}_{T},c)}{\partial\mathbf{x}_{1}},\frac{\partial\varphi(\mathcal{X}_{T},c)}{\partial\mathbf{x}_{2}},...,\frac{\partial\varphi(\mathcal{X}_{T},c)}{\partial\mathbf{x}_{T}}]$
	and the Hessian matrix of $\varphi(\mathcal{X}_{T},c)$ \ac{wrt} $\mathcal{X}_{T}$ is given by
	\begin{align*}
		\mathbf{H}(\mathcal{X}_{T},c) & =\left[\begin{array}{cccc}
			\frac{\partial\varphi(\mathcal{X}_{T},c)}{\partial\mathbf{x}_{1}\partial\mathbf{x}_{1}} & \frac{\partial\varphi(\mathcal{X}_{T},c)}{\partial\mathbf{x}_{1}\partial\mathbf{x}_{2}} & \cdots & \frac{\partial\varphi(\mathcal{X}_{T},c)}{\partial\mathbf{x}_{1}\partial\mathbf{x}_{T}}\\
			\frac{\partial\varphi(\mathcal{X}_{T},c)}{\partial\mathbf{x}_{2}\partial\mathbf{x}_{1}} & \frac{\partial\varphi(\mathcal{X}_{T},c)}{\partial\mathbf{x}_{2}\partial\mathbf{x}_{2}}\\
			\vdots &  & \ddots\\
			\frac{\partial\varphi(\mathcal{X}_{T},c)}{\partial\mathbf{x}_{T}\partial\mathbf{x}_{1}} & \frac{\partial\varphi(\mathcal{X}_{T},c)}{\partial\mathbf{x}_{T}\partial\mathbf{x}_{2}} & \cdots & \frac{\partial\varphi(\mathcal{X}_{T},c)}{\partial\mathbf{x}_{T}\partial\mathbf{x}_{T}}
		\end{array}\right]
	\end{align*}

	Then, the optimal solution is given by Algorithm \ref{alg:initial}. Algorithm \ref{alg:initial} starts with a strictly feasible trajectory $\mathcal{X}^{(0)}$, an initial penalty parameter $c^{(0)}$, along with its weight coefficient $\mu$, as well as the tolerance parameters $\epsilon_{1}$ and $\epsilon_{2}$, which can be selected according to \cite{boyd2004convex}. We then perform alternating optimization on $\mathcal{X}$ and $c$. Specifically,
	we compute $\mathcal{X}^{(m,n)}$ for a sequence of increasing values of $c$, until $c\geq(T-1)/\epsilon_{2}$, which guarantees that we have an $\epsilon_{2}$-suboptimal solution of the original problem. At each iteration, we compute the central point $\mathcal{X}^{(m,n)}$ starting from the previously computed central point, and then increase $c$ by a factor $\mu > 1$. 
	We refer to each centering step as an outer iteration, and Newton’s method is used in the centering step. We refer to the Newton iterations executed during the centering step as inner iterations. 
	
	\section{Numerical Results}
	
	In this section, we begin by introducing the real dataset collection employed in our experiments (Section \ref{sec:data-collect}). We then outline the specifics of the experimental setup (Section \ref{sec:exp-set}). Subsequently, we present the comparison results with the baseline methods (Section \ref{sec:perf}). Finally, we perform a parameter sensitivity analysis (Section \ref{sec:para-sense}) and present the findings of our ablation study (Section \ref{sec:ablation}).

	\subsection{Real Dataset Collection}
	\label{sec:data-collect}
	We collected data in urban areas of Chengdu and Shenzhen, China, where 5G \acpl{bs} are extensively deployed. The layout and antenna configurations of the \acpl{bs} vary between the two cities, alongside differences in geographic environments.
	A HUAWEI Mate 40 Pro smartphone in the vehicle records the \ac{rss} emitted from 5G \acpl{bs} around the vehicle. The data collection process unfolded as follows: 
	
	\begin{enumerate}
		\item[$\bullet$] \textit{Dataset I}: We conducted a data collection procedure in Chengdu, China, on 16th October 2022. This procedure extended over 3.5 hours and covered a total distance of 161 km.
		\item[$\bullet$] \textit{Dataset II}: Another data collection procedure was carried out in Shenzhen, China, starting on the 22nd of April 2023. This procedure lasted for 2.9 hours and encompassed a travel distance of 142 km.
	\end{enumerate}

	The driving test was conducted along a randomly chosen route in both cities. The maximum speed limit in this urban area is $v_{\mathrm{max}} = 22.2$ m/s (equivalent to 80 km/h). We maintained a data collection interval of $\delta = 0.2$ seconds, meaning that the \ac{rss} measurements were recorded at regular intervals along the route. As shown in Figure~\ref{fig:velocity-pdf}, the actual maximum driving speed is $105.7$ km/h and $110.5$ km/h for Dataset I and II, respectively. Ideally, vehicles should not exceed the maximum speed limit, but it's been observed that they often do during actual driving. It's worth noting that when driving in urban areas, exceeding the maximum speed limit by 10\% is not legally punishable. The real average speed is 13.3 m/s (48.2 km/h) for Dataset I and 13.8 m/s (49.8 km/h) for Dataset II.

	Table \ref{tab:per-road} shows the proportion of road types where vehicles traveled across two datasets. Urban roads make up the majority, followed by national and provincial (NP) roads and highways. The least represented are rural roads.
	Datasets I and II provide extensive trajectory data covering diverse driving scenarios and complex environments, characterized by lengthy routes, varied road conditions, and inherently sparse and noisy \ac{rss} signals, presenting significant challenges in accurately determining vehicle trajectories.
	\begin{table}[t]
		
		\caption{Percentage [\%] of road types.}
		\centering{}
		\begin{tabular}{l|cccc}
			\toprule[1.5pt]  
			& \multicolumn{4}{c}{Road types}\tabularnewline
			& Urban road & Highways & NP roads & Rural roads\tabularnewline
			\hline 
			Dataset I & 58.05 \% & 12.24 \%& 21.17 \%& 8.54 \%\tabularnewline
			Dataset II & 46.11 \%& 25.35 \%& 24.11 \%& 4.43 \%\tabularnewline
			\bottomrule[1.5pt]  
		\end{tabular}
		\label{tab:per-road}
	\end{table}
	\begin{table*}[t]
		%	\begin{centering}
			\caption{\ac{vtr} performance comparison of our method and the baselines.}
			%		\par\end{centering}
		%	\vspace{0.1in}
		\centering{}%
		\begin{tabular}{>{\raggedright}p{0.2cm}>{\raggedright}p{1.5cm}|>{\centering}p{1.2cm}>{\centering}p{1.2cm}>{\centering}p{1.2cm}>{\centering}p{1.2cm}>{\centering}p{1.2cm}>{\centering}p{1.2cm}>{\centering}p{1.2cm}>{\centering}p{0.8cm}>{\centering}p{0.8cm}>{\centering}p{0.8cm}}
			\toprule[1.5pt]
			& & MaR~\cite{iqbal2014development} & WCL~\cite{MagGioKanYu:J18} &HMMM~\cite{newson2009hidden}&STMM~\cite{lou2009map}&STDMM~\cite{hsueh2018map}&
			TMM~\cite{jin2022transformer}&AMM~\cite{hu2023amm}& MSR & HRE&HREA\tabularnewline
			\hline 
			\multirow{2}{*}{I} &$QLE$ [m] & 116.6 & 85.2 & 64.3&57.8&52.6	&46.9&	45.3&41.4 & \underline{12.6}&\textit{10.1}\tabularnewline
			%		\hline 
			&$TME$ [\%]  & 74.14 & 48.32 &32.11&28.25	&20.43&	15.97	&10.54& 8.93 & \underline{0.34}&\textit{0.28}\tabularnewline
			\hline
			\multirow{2}{*}{II} &$QLE$ [m] & 125.9 & 88.6 & 67.7&59.3&	55.6	&48.2&	47.7&45.0 & \underline{14.7}&\textit{12.4}\tabularnewline
			%		\hline 
			&$TME$ [\%]  & 76.24 & 51.91 &37.44&28.48&	21.94&	16.37	&11.42& 9.75 & \underline{0.37}&\textit{0.25}\tabularnewline
			\bottomrule[1.5pt]
		\end{tabular}
		\label{tab:loc-traj-performance}
	\end{table*}
	
	\subsection{Experimental Setup}
	\label{sec:exp-set}
	
	\subsubsection{Compared Methods}
	
	We establish three baseline methods for comparison: the Max \ac{rss} (MaR)
	approach \cite{iqbal2014development}, and the \ac{wcl} approach \cite{MagGioKanYu:J18}. In the MaR method, at time $t$,
	the strongest signal among the surrounding $Q$ \acpl{bs} is selected, and
	the estimated position is given by the position of the chosen \ac{bs}.
	The \ac{wcl} method involves finding an approximate weighed location
	$\{\mathbf{z}_{t}\}$ as the same as Section \ref{subsec:Location-Initialization}. MSR is the proposed rough trajectory estimation method in Section \ref{subsec:Location-Initialization}.

	We also compare with some traditional and state-of-the-art map matching methods, including
	Hidden Markov-based map-matching (HMMM)~\cite{newson2009hidden}, 
	map-matching for low-sampling-rate GPS trajectories (STMM)~\cite{lou2009map}, spatio-temporal based map-matching algorithm (STDMM)~\cite{hsueh2018map}, Transformer-based map-matching (TMM)~\cite{jin2022transformer}, and
	adaptive map map-matching algorithm (AMM)~\cite{hu2023amm}. These methods run with the location given by WCL method.
	These map-matching methods recover vehicle trajectories by aligning raw GPS data to the most appropriate path on the road network. In contrast, our study focuses on trajectory map-matching using only RSS measurements. To facilitate a fair comparison with advanced map-matching methods, we employ WCL to estimate locations based on RSS and use these estimated locations as input for the map-matching methods.

	\begin{figure}[t]
		\begin{centering}
			\includegraphics[width=1\columnwidth]{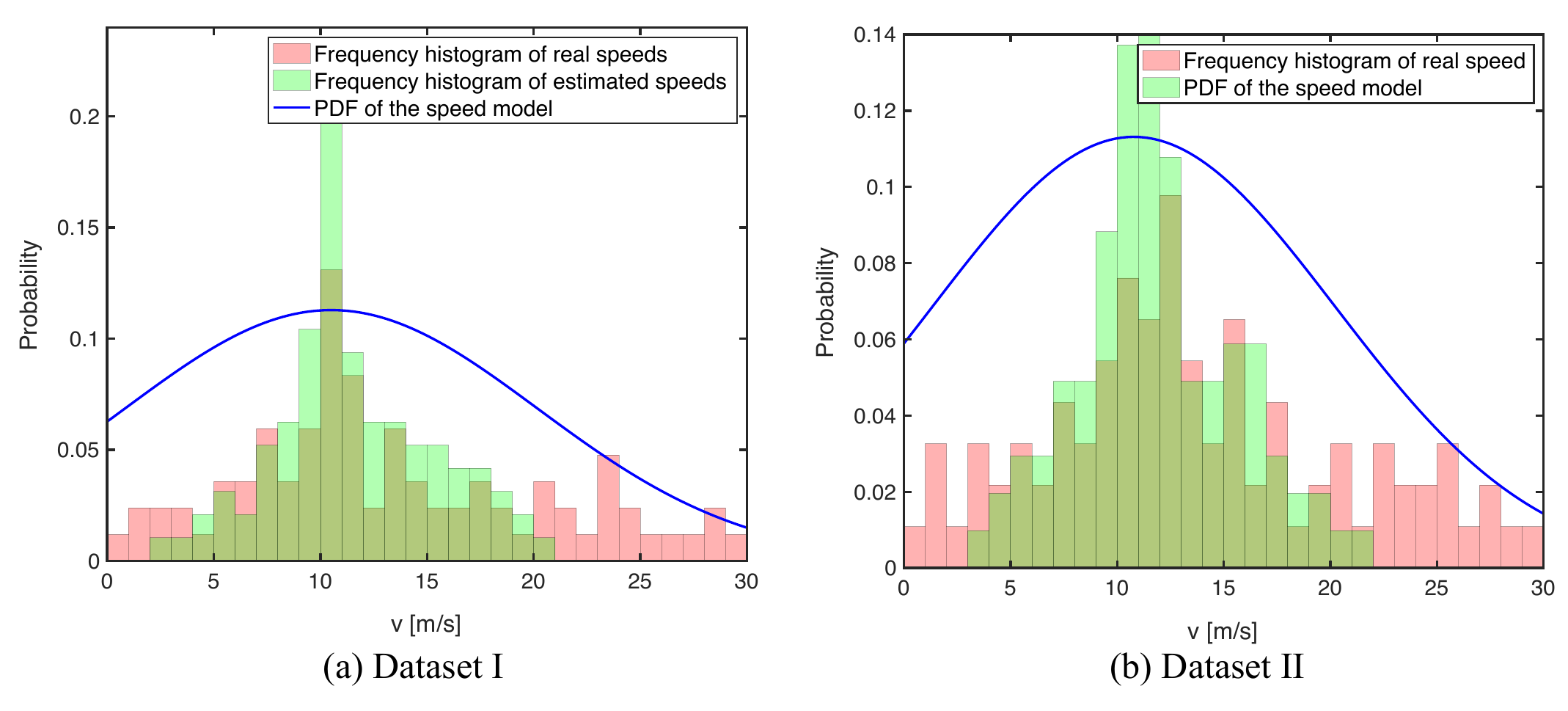}
			\par\end{centering}
		%	\vspace{-0.3in}
		\centering{}
		\caption{{The frequency histogram of real speeds and recovered speeds v.s. the \ac{pdf} of the speed model.}}
		\label{fig:velocity-pdf}
		%		\vspace{-0.1in}
	\end{figure}

	\subsubsection{Evaluation Metrics}
	Queuing location error (QLE) has been used in \cite{chen2022vehicle} and was applied here for better comparison. QLE illustrates the average Root-Mean-Square Error (RMSE) of the queuing status between the reconstructed and ground-truth trajectories: 
	\begin{align}
		\label{eq:qle}% zheng remove label
		QLE=\sqrt{\frac{1}{T}\sum_{t=1}^{T}\|\mathbf{x}_{t}^{*}-\tilde{\mathbf{x}}_{t}\|_{2}^2}
	\end{align}
	where $\tilde{\mathbf{x}}_{t}$ and $\mathbf{x}_{t}^{*}$ are the positions actual reconstruct for the ground-truth and reconstructed trajectories at the same time step. $T$ represents the set of queuing status.
	
	Trajectory
	matching error (TME) has been used in \cite{hu2023amm}, which is defined as 
	\begin{align*}
		\label{eq:tme}
		TME=\frac{L_{loss}+L_{surplus}}{L}\times 100\%
	\end{align*}
	where $L_{loss}$ represents the length of paths included in the trajectory
	but not matched, $L_{surplus}$ is the length of the paths matched
	but not included in the trajectory, and $L$ is the length of the
	ground truth trajectory. We utilize the classical map-matching algorithm \cite{newson2009hidden} to match the position results of the baselines to the most reasonable roads and create a continuous trajectory for the calculation of TME.

	\subsubsection{Parameter Setting}
	For our method, the average speed $v_{\text{avr}}$ is set to be $38$
	km/h (equivalent to $10.5$ m/s) and $39$
	km/h (equivalent to $10.8$ m/s) for Dataset I and II separately. The probability of driving with the maximum limit speed $v_{\mathrm{max}}$ is 
	$80$ km/h (equivalent to 22.2 m/s), and the corresponding probability $\eta=0.05$.  For the proposed adaptive mobility model, we set $A=10$.

	\subsection{Performance Comparison with Fixed Mobility Model}
	\label{sec:perf}
	
	\subsubsection{Performance with Original Data}

	Table \ref{tab:loc-traj-performance} presents the QLE and TME performance of our method. Our HRE exhibits the smallest TME of 0.34\% and a QLE of 12.6 meters when compared to all baseline methods on Dataset I. Similarly, on Dataset II, it achieves the smallest TME of 0.37\% and a QLE of 11.2 meters when compared to all baseline methods.
	Experimental result shows that the localization errors at each time slot i.e., $\|\mathbf{x}_{t}^{*}-\tilde{\mathbf{x}}_{t}\|_{2}$, of our HRE are all within 60 meters and those of our MSR are all within 100 meters, while the localization errors of the baselines may be greater than 300 meters.

	As shown in Figure \ref{fig:velocity-pdf}, the red histogram represents the speed probability statistics calculated from the actual driving trajectory. The green histogram is derived from the recovered  driving trajectory, and the blue line represents the probability distribution of driving speeds from the proposed mobility model. Due to the guidance and constraints of the proposed mobility model, the driving speeds in the recovered  trajectory are predominantly concentrated around the average speed and do not exceed the maximum speed limit.

	As shown in Table \ref{tab:loc-traj-performance}, compared with map-matching methods~\cite{newson2009hidden, lou2009map, hsueh2018map, jin2022transformer, hu2023amm}, the proposed HRE method demonstrates better performance than these map-matching methods. The primary reason for this superior performance is that the proposed approach directly infers vehicle positions on the road network from \ac{rss} measurements. In contrast, the map-matching methods first estimate a rough location based on \ac{rss} using WCL and then match the rough location to a specific position on the road network. This process introduces errors both in the location estimation using WCL and in the map-matching, leading to a "double error" effect. Additionally, the initial location estimation using the location of BSs without considering signal propagation model and vehicle mobility model, which can result in significant inaccuracies.
	In contrast, the proposed HRE method directly leverages the signal propagation of \ac{rss} measurements to search for locations within the road network with the guidance of the mobility model, allowing it to embed the \ac{rss} sequence into a trajectory with higher precision.

	\begin{figure}[t]
		\begin{centering}
			\includegraphics[width=1\columnwidth]{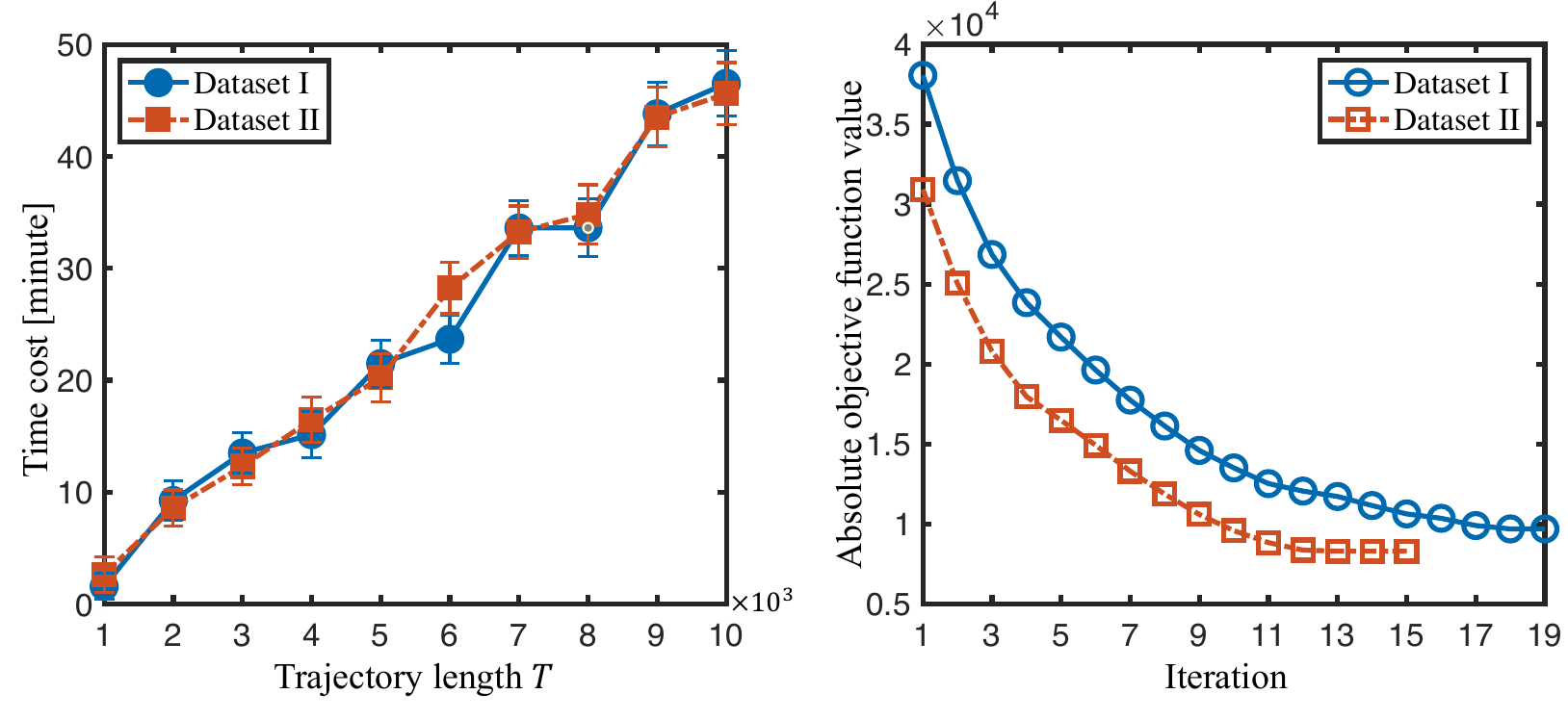}
			\par\end{centering}
		%	\vspace{-0.3in}
		\centering{}\caption{(a) Relationship between trajectory length and the runtime of Algorithm~\ref{alg:overal} on Dataset I. 
			The figure displays the mean values (depicted by points) from 200 runs of randomly generated trajectories with length $T$, accompanied by $\pm 3\sigma$ confidence intervals.
			(b) Convergence behavior of Algorithm~\ref{alg:overal} on Dataset I and II.}
		\label{fig:Effect_time}
		%	\vspace{-0.1in}
	\end{figure}

	As shown in Figure \ref{fig:Effect_time}(a), Algorithm \ref{alg:overal} takes approximately 2 minutes to estimate 1,000 trajectory positions and 45 minutes for 10,000 positions. The proposed Algorithm \ref{alg:overal} exhibits linear complexity, with the runtime increasing by about 5 minutes for every additional 1,000 trajectories.
	In addition, we randomly extract trajectories of length $T$ from Dataset I for recovery, conducting 200 random trials. As shown in Figure~\ref{fig:Effect_time}(a), the performance of Algorithm~\ref{alg:overal} is relatively stable. The runtime exhibits slight fluctuations because the amount of RSS data included in the randomly extracted trajectories varies.
	To address the high data volumes in complex urban networks, we can partition the entire trajectory into smaller segments, each containing 1,000 trajectory positions. Subsequently, Algorithm \ref{alg:overal} can be executed on each segment in parallel, significantly reducing the total computation time.
	As demonstrated in Figure~\ref{fig:Effect_time}(b), the proposed Algorithm~\ref{alg:overal} converges within 19 iterations.
	
	\begin{table}[t]
		%	\begin{centering}
			\caption{Performance (QLE [m]) of the proposed method on different road types.}
			\centering{}
			\begin{tabular}{cc|cccc}
				\toprule[1.5pt]   
				\multirow{2}{*}{Dataset} & \multirow{2}{*}{Method} & \multicolumn{4}{c}{Road types}\tabularnewline
				&  & Urban road & Highways & NP roads & Rural roads\tabularnewline
				\hline 
				\multirow{2}{*}{ I} & HRE & 17.4 m & 9.5 m & 11.6 m & 19.6 m\tabularnewline
				& HREA & 10.3 m & 9.2 m & 9.7 m & 10.8 m\tabularnewline
				\hline 
				\multirow{2}{*}{ II} & HRE & 18.3 m & 10.7 m & 12.1 m & 19.7 m\tabularnewline
				& HREA & 12.6 m & 11.4 m & 11.8 m & 12.8 m\tabularnewline
				\bottomrule[1.5pt]   
			\end{tabular}
			\label{tab:perfor-road}
		\end{table}

		\begin{table}[t]
			%	\begin{centering}
				\caption{Performance (QLE [m]) of the proposed method across different traffic volumes.}
				\centering{}
				\begin{tabular}{cc|ccc}
					\toprule[1.5pt]   
					\multirow{2}{*}{Dataset} & \multirow{2}{*}{Method} & \multicolumn{3}{c}{Vehicle density}\tabularnewline
					&  & Low & Medium & High\tabularnewline
					\hline 
					\multirow{2}{*}{Dataset I} & HRE & 10.4 m & 12.8 m & 15.8 m\tabularnewline
					& HREA & 9.6 m & 10.2 m & 10.5 m\tabularnewline
					\hline 
					\multirow{2}{*}{Dataset II} & HRE & 12.5 m & 15.2 m & 17.1 m\tabularnewline
					& HREA & 12.1 m & 12.5 m & 12.9 m\tabularnewline
					\bottomrule[1.5pt]   
				\end{tabular}
				\label{tab:perfor-density}
			\end{table}

			To explore the impact of road types and vehicle density on the performance of the proposed method, we analyzed the VTR errors on different types of roads. As shown in Table \ref{tab:perfor-road}, vehicles traveling on highways exhibit the lowest errors due to relatively stable driving speeds, making VTR recovery more straightforward. Conversely, vehicles on rural roads display higher errors due to greater speed variability, complicating VTR accuracy.

			As illustrated in Table \ref{tab:perfor-density}, we categorized the traffic flow on roads into three density levels: low, medium, and high. Vehicles in areas with low vehicle density have smaller VTR errors because they can maintain more stable speeds, facilitating easier VTR recovery. However, in high-density conditions, frequent speed changes increase VTR errors significantly.

			\subsubsection{Performance with Sporadic Data}
\begin{figure}[t]
				\begin{centering}
					\includegraphics[width=1\columnwidth]{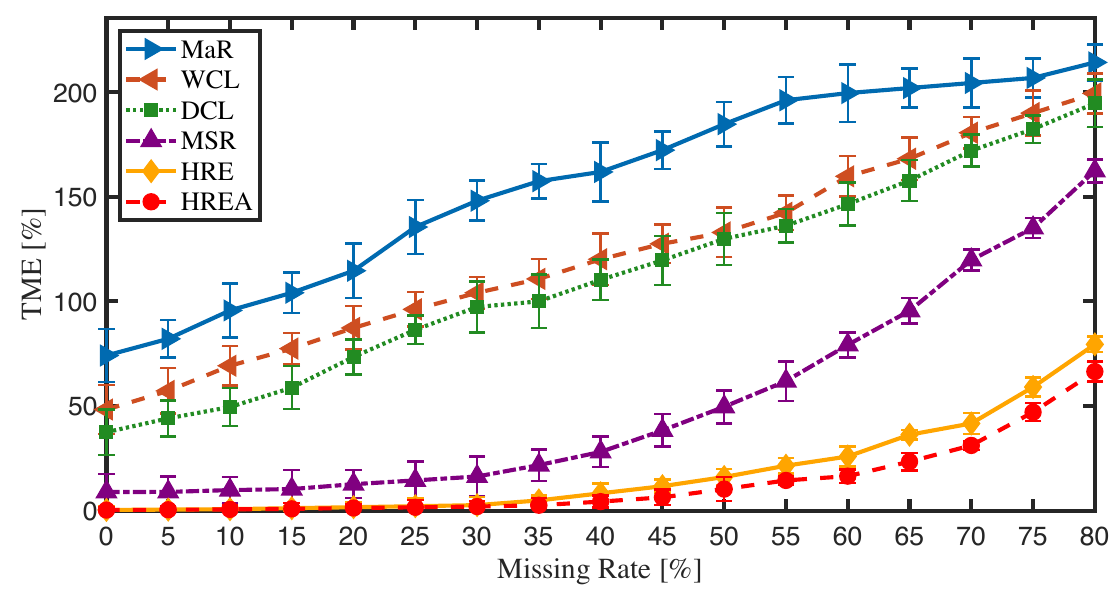}
					\par\end{centering}
				%	\vspace{-0.3in}
				\centering{}\caption{ 
					Relationship between TME performance and missing rate on Dataset I.
					The figure displays the mean values (depicted by points) from 200 runs 
					with data missing at each missing rate, accompanied by $\pm 3\sigma$ confidence intervals.
					\label{fig:MissCurve-irregu}}
				%	\vspace{-0.21in}
			\end{figure}

			We simulate varying degrees of \ac{rss} sporadic characteristics by introducing random data missing in the dataset, aiming to validate the robustness of our algorithm in handling different levels of \ac{rss} sporadic patterns.
			Figure \ref{fig:MissCurve-irregu} illustrates the TME performance of our method on Datasets I and II at various missing rates. Here, a missing rate of 30\% indicates that 30\% of \ac{rss} values were randomly selected and removed from the dataset. As the missing data rate increases, our MSR and HRE exhibit minimal performance degradation compared to the baselines. It is noteworthy that the TME of all baselines exceeds 100\% when the missing rate is 30\%. This implies that these methods predict a significant number of erroneous trajectories. The proposed method maintains relatively stable performance with up to 30\% data loss. However, as the data loss rate increases beyond this point, the performance degrades significantly. This degradation occurs because, at each time step, the user can only receive RSS values from one primary cell and six neighboring cells, resulting in a total of seven RSS measurements. For accurate RSS-based localization, signals from at least three BSs are required. When a substantial portion of the data is missing, the number of available RSS measurements for trajectory recovery often falls below three, leading to large estimation errors under high data loss rates. When the data missing rate reaches 70\%, the TME performance of the proposed HRE and HREA achieves 50\%, and only 0--3 RSS values are sampled at each time slot, greatly increasing the difficulty of trajectory recovery. As the missing rate continues to increase, the error of the proposed method increases sharply because a higher missing rate means that RSS data may not be collected at every moment. In addition, we randomly introduce missing mask 200 times. As shown in Figure \ref{fig:MissCurve-irregu}, the performance of HRE and HREA is relatively stable. The TME performance exhibits slight fluctuations because RSS data with inherently large deviations are deleted or retained, leading to reduced or increased errors \cite{feng2024docpedia,zhao2024harmonizing,wang2024pargo,sun2024attentive,lu2024bounding,zhao2024tabpedia,tang2024mtvqa,tang2024textsquare,shan2024mctbench,feng2023unidoc,xu2024textit}.

\begin{figure}[t]
				\centering{}\includegraphics[width=1\columnwidth]{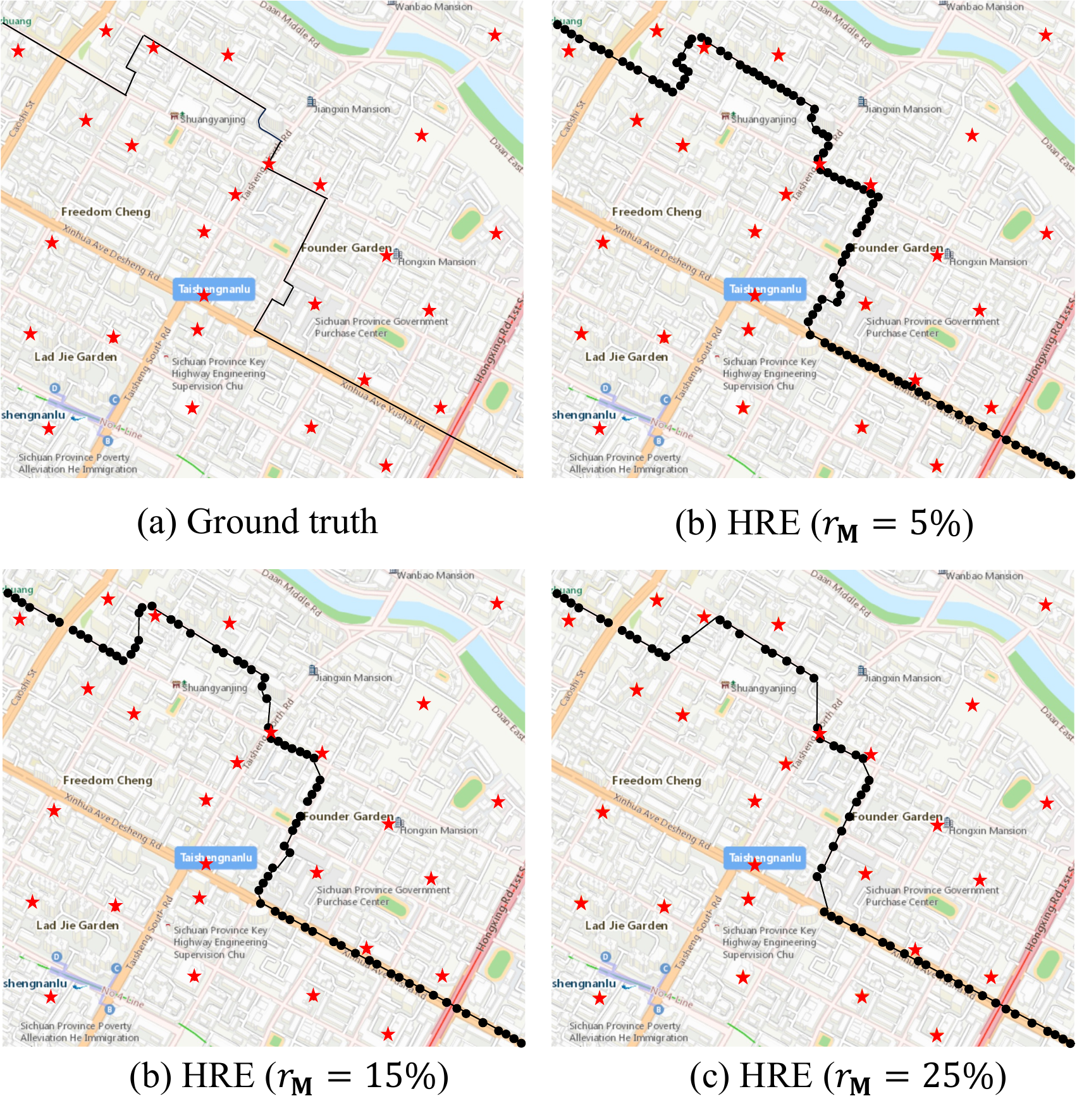}
				\caption{Example trajectory estimation result of our HRE with different missing rates $r_{\text{M}}$. The red pentagons represent \acpl{bs}, while the black dots indicate the estimated vehicle positions using our HRE method. \label{fig:MissTraj}}
				%	\vspace{-0.1in}
			\end{figure}

			It's worth noting that our HRE can still estimate an approximate location in the event that no \ac{rss} is measured at a specific time. Specifically, if at time $t$ the vehicle does not measure any \ac{rss}, there will be no emission probability at that time slot, but transition probabilities will be available. Our HRE will estimate the position at time slot $t$ based on the \ac{rss} measurements at time slots $t-1$ and $t+1$.
			Figure~\ref{fig:MissTraj} demonstrates the performance of our HRE on a short segment of the path in the presence of sporadic data. Despite the escalating data missing rate, our algorithm continues to successfully recover the trajectory of the vehicle, albeit with some details of the journey being overlooked.

			% zheng 变成一个图，横坐标加到60，加入曲线HREA

			\subsection{Performance Comparison with Adaptive Mobility Model}
			\label{sec:perf2}
			
			As shown in Table \ref{tab:loc-traj-performance}, our HREA exhibits the smallest TME of 0.28\% and a QLE of 10.1 meters when compared to all baseline methods and the proposed HRE on Dataset I. Similarly, on Dataset II, it achieves the smallest TME of 0.25\% and a QLE of 12.4 meters when compared to all baseline methods and the proposed HRE.

			As shown in Table \ref{tab:perfor-road}, although the proposed HREA exhibits varying performance across different road types, the performance differences of HREA across these road types are minimal compared to HRE. In Dataset I, the performance gap between the maximum QLE (on rural roads) and the minimum QLE (on highways) is only 1.6\,m for HREA, whereas for HRE, this gap is 10.1\,m. In Dataset II, the gap is 9.0\,m for HRE and 1.4\,m for HREA.

			As illustrated in Table \ref{tab:perfor-density}, although the proposed HREA shows varying performance across different vehicle densities, the performance differences of HREA across these densities are minimal compared to HRE. In Dataset I, the performance gap between the maximum QLE (high density) and the minimum QLE (low density) is only 0.9\,m for HREA, whereas for HRE, this gap is 5.4\,m. In Dataset II, the gap is 4.6\,m for HRE and 0.8\,m for HREA.

			As shown in Figure \ref{fig:MissCurve-irregu}, as the missing data rate increases, the TME performance of the proposed HREA gradually deteriorates, similar to the proposed HRE. However, the proposed HREA consistently outperforms HRE across all missing data rates.

			\subsection{Parameter Sensitivity Analysis}
			\label{sec:para-sense}
			
			\subsubsection{Effect of the parameters $v_{\text{avr}}$, $\eta$, and $v_{\mathrm{max}}$ in fixed mobility model} 
\begin{figure}[t]
				\begin{centering}
					\includegraphics[width=1\columnwidth]{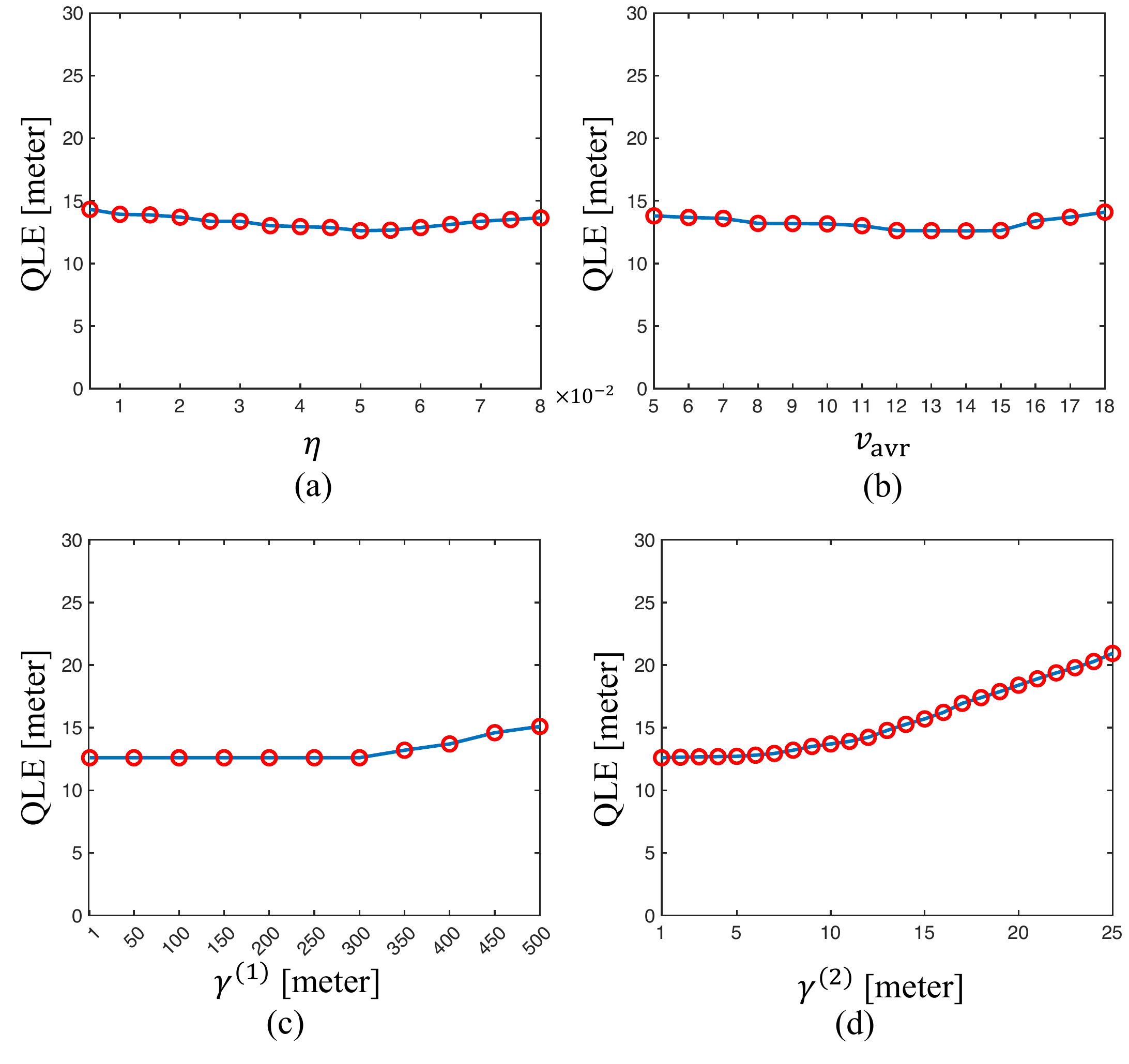}
					\par\end{centering}
				%	\vspace{-0.3in}
				\centering{}\caption{The relationship between the QLE performance of our HRE and the parameters (a) $v_{\text{mean}}$, and (b) $\eta$ on Dataset I. The error $QLE$ v.s. (c) the space $\gamma^{(2)}$, and (d) the space $\gamma^{(2)}$ on Dataset I.}
				\label{fig:ParaSens_mobility}
				%		\vspace{-0.1in}
			\end{figure}

			The maximum speed limit $v_{\mathrm{max}}$ can be readily obtained from publicly available information specific to the region. However, determining the exact $v_{\text{avr}}$ and $\eta$ is challenging. This section assesses the sensitivity of our HRE to $v_{\text{avr}}$ and $\eta$.
			Keeping other parameters fixed, we varied the $\eta$ value. The QLE performance of our HRE algorithm on Dataset I is shown in Figure \ref{fig:ParaSens_mobility} (a). Our algorithm remains stable for $0.001 \leq \eta \leq 0.08$.
			Similarly, to validate the robustness of our HRE algorithm with respect to $v_{\text{avr}}$ values, we conducted the following experiments. While keeping other parameters constant, we varied the $v_{\text{avr}}$ value and observed the performance of our HRE algorithm on Dataset I. Our algorithm remains stable for $5 \leq v_{\text{avr}} \leq 18$ ($0.2v_{max} \leq v_{\text{avr}} \leq 0.8v_{max}$) as shown in Figure \ref{fig:ParaSens_mobility} (b).
			
			\begin{figure}[t]
				\begin{centering}
					\includegraphics[width=1\columnwidth]{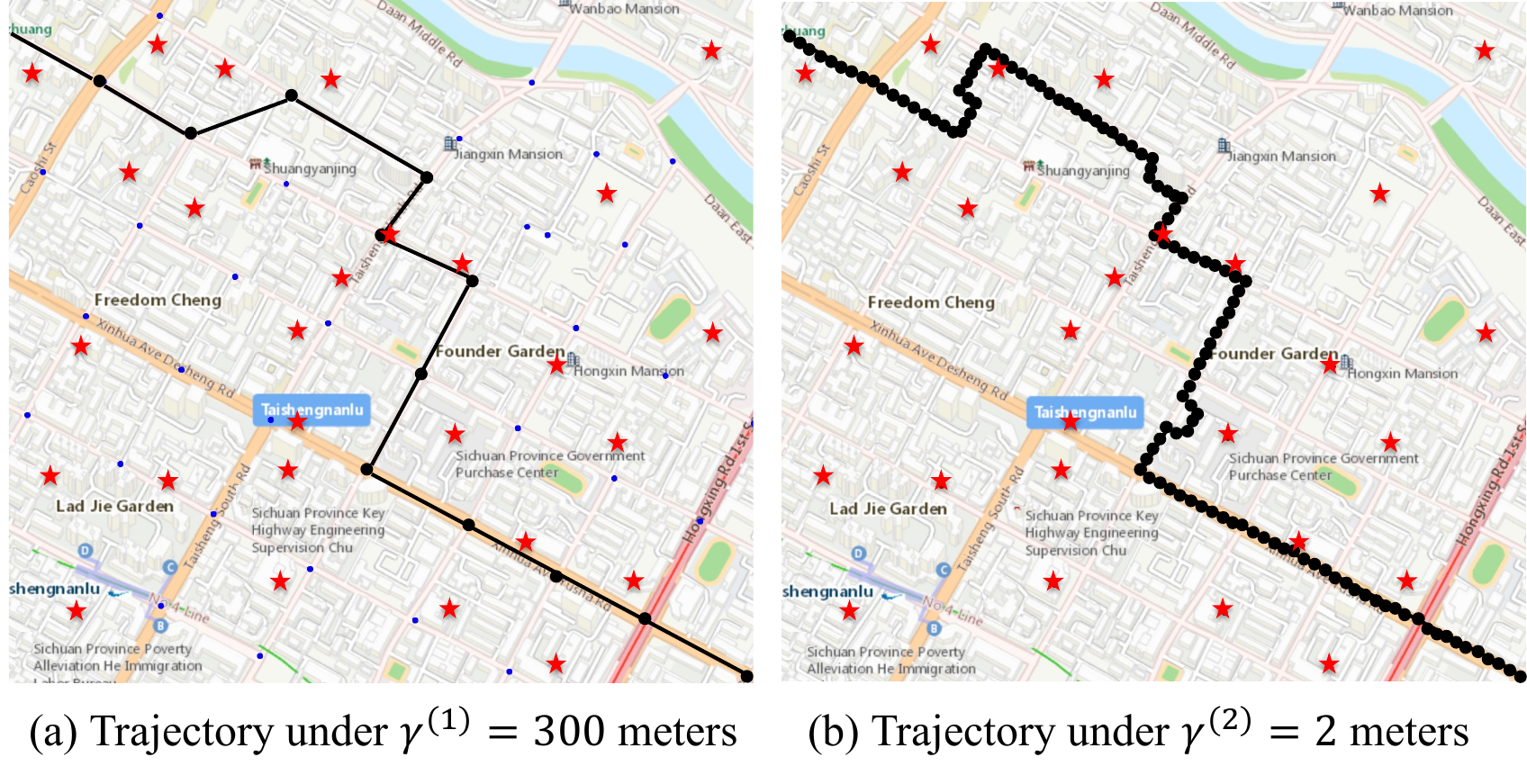}
					\par\end{centering}
				%	\vspace{-0.3in}
				\centering{}\caption{Example procedure of Algorithm \ref{alg:viterbi} with  $\gamma^{(1)}=300, \gamma^{(2)}=2$. The blue dots represent nodes in the road graph.}
				\label{fig:Gamma_comp}
				%	\vspace{-0.1in}
			\end{figure}
			
			\begin{table}[t]
				\caption{Comparison of the proposed HRE with different initialization strategies on the Dataset I.}
				\centering{}%
					\begin{tabular}{>{\raggedright}p{1.4cm}|>{\centering}p{0.7cm}>{\centering}p{0.7cm}>{\centering}p{0.7cm}>{\centering}p{0.9cm}>{\centering}p{0.7cm}>{\centering}p{0.9cm}}
						\toprule[1.5pt] 
						\textit{} & MaR & WCL  & $\Theta$ (R) &   $\mathcal{X}_T$ (R)& MSR &$\Theta$ (GT) \tabularnewline
						\hline 
						$QLE$ [m]& 38.4 & 30.5  & 47.8 & 74.8& 12.6& 8.4  \tabularnewline
						$TME$ [\%]  & 27.24 & 19.54  & 35.22 & 28.18& 0.34& 0.27 \tabularnewline
				
						\bottomrule[1.5pt]
					\end{tabular}
					\label{tab:ablation-MSR}
				\end{table}

				\begin{table}[t]
					\caption{Comparison of different map-matching methods with the result of MSR on Dataset I.}
					\centering{}%
					%	\begin{tabular}{l|cccccc}
						\begin{tabular}{>{\raggedright}p{1.4cm}|>{\centering}p{0.7cm}>{\centering}p{0.7cm}>{\centering}p{0.7cm}>{\centering}p{0.9cm}>{\centering}p{0.7cm}>{\centering}p{0.8cm}}
							\toprule[1.5pt] 
							\textit{} & HMMM & STMM  & STDMM&  TMM& AMM &HRE \tabularnewline
							\hline 
							$QLE$ [m]& 40.1	&38.1	&35.5	&30.3	&24.3&	12.6  \tabularnewline
							$TME$ [\%]  & 17.37	&13.52	&10.33	&6.57	&4.13	&0.34 \tabularnewline
				
							\bottomrule[1.5pt]
						\end{tabular}
						\label{tab:ablation-HRE}
					\end{table}

					\subsubsection{Effect of equal spacing $\gamma^{(2)}$ and $\gamma^{(2)}$ in the construction of the mobility graph} 
					Figure \ref{fig:Gamma_comp} illustrates the two-stage procedure of our Algorithm \ref{alg:viterbi}. Firstly, based on $\gamma^{(1)}$, we select equal spacing states within the road network and construct a corresponding coarse-grained graph. Subsequently, the algorithm is executed to obtain an approximate trajectory. Using this approximate trajectory, we identify the road segments that the vehicle may have traversed. Then, on these potential road segments, we employ $\gamma^{(2)}$ to select equal spacing states and create the corresponding fine-grained graph. Finally, on the fine-grained graph, we determine the optimal path that the vehicle actually traveled. The choice of parameter $\gamma^{(1)}$ is typically a relatively large value, e.g., 300 meters. The selection of the value for $\gamma^{(2)}$ plays a crucial role in determining the algorithm's final output quality.

					Figure \ref{fig:ParaSens_mobility} (c) illustrates the relationship between the parameter $\gamma^{(1)}$ and the localization performance of our method. Our algorithm is not particularly sensitive to $\gamma^{(1)}$. However, when $\gamma^{(1)}$ is large, it may lead to a significant deviation in the estimated trajectory based on $\gamma^{(1)}$, resulting in a slight decrease in the performance of our HRE.
					Figure \ref{fig:ParaSens_mobility} (d)
					shows the relationship between the parameter $\gamma^{(2)}$ and the
					localization performance of our method. As the value of $\gamma^{(2)}$
					increases, more states within the set $\mathcal{V}$ become available,
					leading to a more accurate estimated position.

					\subsection{Ablation Study}
					\label{sec:ablation}
					
					First, we explore six baseline initialization methods, including MaR, WCL, random propagation model parameter \(\bm{\Theta}\), random trajectory, and the ground truth (GT) propagation model parameter $\Theta$. The comparison results are summarized in Table \ref{tab:ablation-MSR}. Initializing HRE with our MSR method leads to even lower errors, underscoring the value of MSR. It is worth noting that random initialization of the propagation model parameters or trajectory can result in significantly larger errors, as the random initialization may lead the iterative optimization algorithm to converge to a poor local solution. If the parameters of the signal propagation model are known, the trajectory can be directly determined using Viterbi algorithm, thereby eliminating the need for iterative optimization. In this case, the error of trajectory recovery is minimized for the proposed method, with values of 8.4 for QLE and 0.27 for TME. However, the proposed HRE method, initialized using MSR, closely approximates this performance even without knowledge of the signal propagation model parameters.
					
					Map-matching methods align a sequence of observed user positions with the road network on a digital map. We utilize the results from MSR as inputs for these map-matching methods. Table \ref{tab:ablation-HRE} illustrates the performance of various map-matching approaches, revealing that all methods underperform compared to the proposed HRE method when initialized with MSR.

					\subsection{Effect of Signal Complexities}
					
					\begin{figure}[t]
						\begin{centering}
							\includegraphics[width=1\columnwidth]{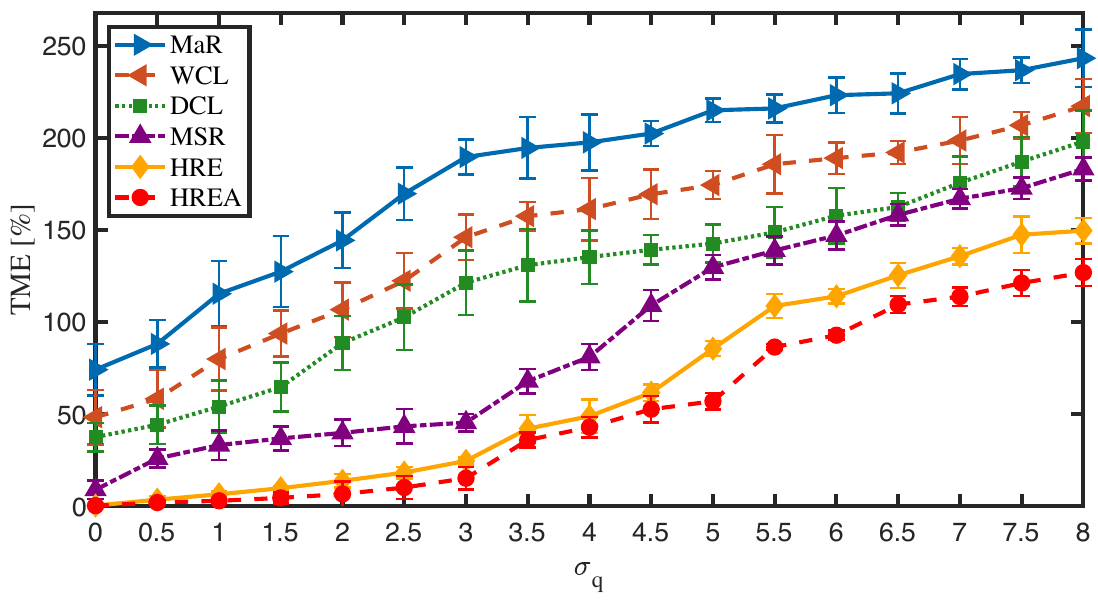}
							\par\end{centering}
						%	\vspace{-0.3in}
						\centering{}\caption{Relationship between noise level and the Performance of the proposed method and comparisons on Dataset I. The figure displays the mean values (depicted by points) from 200 runs on randomly generated noise data with a standard deviation of $\sigma_q$, accompanied by $\pm 3\sigma$ confidence intervals.}
						\label{fig:Effect_sigmaq}
						%	\vspace{-0.1in}
					\end{figure}
					We add zero-mean Gaussian noise \( N(0, \sigma_q^2) \) to the RSS data. Note the practical implications of RSS signal attenuation in real-world scenarios: an increase of 100 meters in the distance from the BS results in an RSS attenuation of approximately 1.24 dB. When the noise variance is set to \( \sigma_q = 3 \), approximately 31.7\% of the RSS measurements experience fluctuations of \(\pm3\) to \(\pm9\) dB. Such significant variations make RSS-based position estimation highly challenging, thus modeling complex signal interactions.

					As illustrated in Figure~\ref{fig:Effect_sigmaq}, we observe that as the noise level increases, the performance of our proposed algorithms, HRE and HREA, deteriorates significantly. Specifically, when the noise level is \( \sigma_q < 3 \), both HRE and HREA maintain a TME of approximately 25\%. In contrast, the comparison methods experience a TME exceeding 120\%, indicating substantial deviations in the estimated trajectories under the same noise conditions. Moreover, we randomly add Gaussian noise \( N(0, \sigma_q^2) \) to Dataset I for 200 trials. As shown in Figure~\ref{fig:Effect_sigmaq}, the performance of HRE and HREA is relatively stable. The TME performance exhibits slight fluctuations because some RSS data already have measurement biases. Continuing to add noise to such RSS data causes the RSS measurements to deviate significantly from their true values, resulting in incorrect position estimations.

					\section{Conclusion}

					In this paper, we introduce two \ac{rss}-based paradigms for \ac{vtr}. The proposed paradigms effectively leverage the hidden spatial-temporal correlation within the sequential, noisy, and sporadic \ac{rss} data while searching for a continuous path within the highly intricate road network. Through empirical validation using two real-world datasets, our proposed method demonstrates outstanding performance, even in scenarios with a high rate of missing data.

					\ifCLASSOPTIONcaptionsoff
					\newpage
					\fi

					
					%{\small
						\bibliographystyle{IEEEtran}
						\bibliography{my_ref}
						%}
			
				\end{document}